\pdfoutput=1
\immediate\write18{command}

\documentclass[sigconf,screen]{acmart}

\copyrightyear{2024} 
\acmYear{2024} 
\setcopyright{acmlicensed}\acmConference[ASE '24]{39th IEEE/ACM
International Conference on Automated Software Engineering }{October
27-November 1, 2024}{Sacramento, CA, USA}
\acmBooktitle{39th IEEE/ACM International Conference on Automated Software
Engineering (ASE '24), October 27-November 1, 2024, Sacramento, CA, USA}
\acmDOI{10.1145/3691620.3695552}
\acmISBN{979-8-4007-1248-7/24/10}

\usepackage{tcolorbox}
\usepackage{hyperref}
\usepackage{appendix}
\usepackage{multirow}
\usepackage{booktabs}
\usepackage{pifont}
\usepackage{makecell}
\usepackage{amsmath}
\usepackage{subcaption}
\usepackage{graphicx}
\usepackage[inkscapelatex=false]{svg}
\usepackage{xcolor}
\usepackage{colortbl}

\newcommand{\yun}[1]{\textcolor{black}{#1}}
\newcommand{\wcz}[1]{\textcolor{black}{{#1}}}

\usepackage[compact]{titlesec}
\titlespacing*{\section}{0pt}{*0.8}{*0.8}
\titlespacing*{\subsection}{0pt}{*0.6}{*0.6}

\newcommand{\finding}[2]{
\begin{tcolorbox}[width=\linewidth,boxrule=0pt,top=0.5pt, bottom=0.5pt,
left=0.5pt,right=0.5pt, colback=blue!5,colframe=blue!5,
before skip=2.3pt, after skip=2.3pt]
\textbf{Finding #1:} %\textit
{#2}
\end{tcolorbox}}

\AtBeginDocument{%
  \providecommand\BibTeX{{%
    \normalfont B\kern-0.5em{\scshape i\kern-0.25em b}\kern-0.8em\TeX}}}

\begin{document}

\title{ComplexCodeEval: A Benchmark for Evaluating Large Code Models on More Complex Code}
% context-dependent

\author{Jia Feng}
\affiliation{%
  \institution{University of Electronic Science and Technology of China}
  \city{Shenzhen}
  \country{China}
  }
\email{jfeng@std.uestc.edu.cn}

\author{Jiachen Liu}
\affiliation{%
  \institution{Harbin Institute of Technology}
  \city{Shenzhen}
  \country{China}
  }
\email{200110207@stu.hit.edu.cn}

\author{Cuiyun Gao}
\authornote{Corresponding author. The author is also affiliated with Peng Cheng Laboratory. }
\affiliation{%
  \institution{Harbin Institute of Technology}
  \city{Shenzhen}
  \country{China}}
\email{gaocuiyun@hit.edu.cn}

\author{Chun Yong Chong}
\affiliation{%
  \institution{Huawei, Hong Kong}
  \city{Hong Kong}
  \country{China}}
\email{chunyong@ieee.org}

\author{Chaozheng Wang}
\affiliation{%
  \institution{The Chinese University of Hong Kong}
  \city{Hong Kong}
  \country{China}}
\email{adf111178@gmail.com}

\author{Shan Gao}
\affiliation{%
  \institution{Huawei}
  \city{Shenzhen}
  \country{China}
  }
\email{gaoshan17@huawei.com}

\author{Xin Xia}
\affiliation{%
  \institution{Huawei}
  \city{Shenzhen}
  \country{China}
  }
\email{xin.xia@monash.edu}

\begin{abstract}
    In recent years, with the widespread attention of academia and industry on the application of large language models (LLMs) to code-related tasks, an increasing number of large code models (LCMs) have been proposed and corresponding evaluation benchmarks have continually emerged. Although existing evaluation benchmarks are helpful for comparing different LCMs, they may not reflect the performance of LCMs in various development scenarios. Specifically, they might evaluate model performance in only one type of scenario (e.g., code generation or code completion), whereas real development contexts are diverse and may involve multiple tasks such as code generation, code completion, API recommendation, and test function generation. Additionally, the questions may not originate from actual development practices, failing to capture the programming challenges faced by developers during the development process.

To address the aforementioned issues, we propose ComplexCodeEval, a new benchmark for evaluating the performance of LCMs in various development scenarios. ComplexCodeEval includes 3,897 Java samples from 1,055 high-star GitHub repositories and 7,184 Python samples from 2,107 high-star repositories. Each function sample in ComplexCodeEval contains multiple annotations (e.g., function signatures, docstrings and reference APIs) to accommodate various downstream tasks. Furthermore, to better reflect diverse development scenarios, each function sample is required to originate from a repository that depends on at least one selected library (based on popularity), and each function sample must invoke at least one API from the selected library. Additionally, each function sample has multiple timestamps to avoid data leakage. Based on ComplexCodeEval, we evaluate the performance of ten LCMs across four tasks (i.e., code generation, code completion, API recommendation, and test case generation) to explore their performance in complex development environments. Furthermore, we conduct an in-depth analysis of the impact of context and data leakage on model performance. Our experimental results reveal several key findings. For instance, LCMs exhibit varying performance across different coding tasks. Additionally, rich contextual information can greatly enhance the performance of LCMs. Moreover, using leaked data for evaluation may lead to an overestimation of model performance, resulting in inaccurate evaluation outcomes that deviate from the performance in practice.

\end{abstract}

\maketitle

\section{Introduction}

Code has become an important application area for LLMs \cite{allal2023santacoder,austin2021program,chen2021evaluating,guo2024deepseek,li2022competition,li2023starcoder,lozhkov2024starcoder,luo2023wizardcoder,ridnik2024code,roziere2023code,wei2023magicoder,zhong2023codegen}, leading to the emergence of numerous LCMs such as Codex \cite{chen2021evaluating}, Copilot \cite{github_copilot}, and CodeLlama \cite{roziere2023code}. LCMs have been widely adopted for various tasks, including code generation, code completion, test case generation, and API recommendation. This widespread adoption of LCMs has remarkably advanced practical development by not only automating repetitive tasks, but also improving the quality of code and accelerating the overall software development process.
% with the emergence of a large number of Large Code Models (LCMs) and they have been widely used in multiple tasks such as code generation, code completion, API recommendation, and test case generation, effectively promoting practical development. 

In order to understand the performance of LCMs on code, researchers have put a lot of effort into building evaluation benchmarks automatically or manually \cite{austin2021program,cassano2023multipl,chandel2022training,chen2021evaluating,haluptzok2022language,hao2022aixbench,hendrycks2021measuring,lai2023ds,wang2022language,wang2022execution,li2024evocodebench,jain2024livecodebench,yu2024codereval,du2023classeval}.
% % These evaluation benchmarks, such as
For instance, HumanEval \cite{chen2021evaluating} and MBPP \cite{austin2021program} can effectively reflect the capabilities of LCMs in code generation. CrossCodeEval \cite{ding2023crosscodeeval} assesses LCMs' performance in cross-file code completion, while CoderEval \cite{yu2024codereval} and EvoCodeBench \cite{li2024evocodebench} evaluate the performance of LCMs in repository-level code generation. These benchmarks provide crucial references and guidance for the improvement and optimization of LCMs.
% , CrossCodeEval can evaluate the performance of LCMs in cross-file code completion, CoderEval and EvoCodeBench can evaluate the performance of LCMs in repository-level code generation,providing important references and guidance for the improvement and optimization of LCMs.

% Researchers have invested greate effort into constructing evaluation benchmarks, either automatically or manually, to comprehend the performance of LCMs on these tasks. Notable examples include HumanEval and MBPP, which effectively gauge the capabilities of LCMs in code generation. Additionally, CrossCodeEval assesses LCMs' performance in cross-file code completion, while CoderEval and EvoCodeBench evaluate their performance in repository-level code generation. These benchmarks provide essential references and guidance for enhancing and optimizing LCMs.

% Although these evaluation benchmarks can effectively reflect the performance of LCMs in code generation, they may not reflect the programming difficulties faced by developers in actual development.
Although these evaluation benchmarks can effectively measure the performance of LCMs in some aspects, they may not accurately represent the programming challenges faced by developers in various development scenarios. Firstly, practical programming is diverse, whereas these benchmarks are evaluated on only one task (e.g., HumanEval focuses on code generation and CrossCodeEval focuses on code completion). Secondly, the samples used in these benchmarks may be manually crafted or sourced from a limited number of code repositories (e.g., EvoCodeBench from only 25 repositories), thereby covering only a narrow range of application domains and failing to effectively represent the challenges encountered in software development. Finally, these benchmarks may risk data leakage, as their samples could have been included in the training data (e.g., the samples of Concode \cite{iyer2018mapping} are sourced from GitHub repositories, and do not address data leakage concerns), potentially distorting the evaluation results.

% the programming in actual development is diverse, while these evaluation benchmarks only focus on the scenario of code generation. Secondly, the samples of these evaluation benchmarks may be manually written or only come from a small number of code repositories, covering a very limited fields, and cannot effectively reflect the difficulties faced in actual development. Finally, these evaluation benchmarks may have the risk of data leakage, because the samples of the evaluation benchmarks may have been included in the training data, leading to distortion of the evaluation results.

\textbf{Benchmark ComplexCodeEval.} To address these limitations, we propose ComplexCodeEval in this paper. ComplexCodeEval is an evaluation benchmark designed to accommodate multiple downstream tasks, accurately reflect different programming environments, and deliberately avoid data leakage issues.
% To address these limitations, in this paper, we propose ComplexCodeEval, which is an evaluation benchmark suitable for multiple downstream tasks, capable of reflecting the real programming environment, and flexibly avoiding data leakage problems. 
ComplexCodeEval includes 3,897 Java samples from 1,055 code repositories and 7,184 Python samples from 2,107 code repositories.
% In order to make ComplexCodeEval more close to the real development scenario,
To ensure that ComplexCodeEval closely mirrors real-world development scenarios, we first screen 69 popular Java third-party frameworks and 55 popular Python third-party packages based on their SourceRank from Libraries.io \cite{librariesio}. % and their relevance to real-world development needs as identified through developer surveys.
These frameworks and packages cover a wide range of fields, such as web development, network communication, data processing and persistence, and security and encryption.
% which cover a wide range of fields, such as web development and network communication, data processing and persistence, security and encryption.
Then, we select high-star repositories on GitHub that depend on these libraries and analyze them to track the usage of each library’s API. Based on API usage frequency, we extract functions that rely on high-frequency APIs from these repositories as samples.
% we select high-star code repositories that depend on these libraries from GitHub, and analyze these code repositories, tracking the usage of each library's API. Based on the usage frequency of APIs, we extract functions that rely on high-frequency APIs from code repositories as samples. 
To ensure ComplexCodeEval's suitability for multiple downstream tasks, we include various annotations for each sample, such as test cases, reference APIs, and docstrings. To avoid data leakage issues, we incorporate multiple timestamps for each function sample, including project creation time, file creation time, and function update time.
% According to the usage frequency of API, we extracted functions that depend on high-frequency API from code repositories as samples. In order to make ComplexCodeEval suitable for multiple downstream tasks, we extracted multiple annotations for each sample, such as test case, reference API and docstring. In order to flexibly avoid data leakage problems, we added multiple time information for each sample, such as project creation time, file creation time and function update time.

\textbf{Empirical study.} Based on ComplexCodeEval, we evaluate ten popular LCMs, including three families of open-source models (i.e., StarCoder2, CodeLlama, and Deepseek-Coder) with different sizes, as well as one closed-source model (i.e., GPT-3.5-Turbo). To assess the performance of LCMs, we evaluate them across four key tasks: code generation, code completion, API recommendation, and test case generation, respectively. We introduce various contextual information—such as file context and function dependencies—into the prompts to examine the impact of different context conditions on LCMs performance.
% To examine the impact of different context conditions, we introduce various context information (such as file context and function dependencies) in the prompts for evaluating LCM performance. 
To address the data leakage issue, we utilize file creation time as the basis for sample division, selecting samples from different timestamps to assess model performance. These experiments allow us to comprehensively understand the performance of LCMs across different tasks and context conditions, and evaluate the impact of data leakage on model performance.
% In order to explore the performance of LCMs on different tasks, we evaluated their performance on four tasks (i.e., code generation, code completion, API recommendation, test case generation). In order to explore the performance of LCMs under different context conditions, we introduced different context information (such as file context, dependencies) in the prompt to evaluate the performance of LCMs. In order to explore the data leakage problem, we used the file creation time as the basis for sample division, and selected samples from different time points to evaluate the performance of the model. Through these experiments, we can comprehensively understand the performance of LCMs under different tasks and different context conditions, and evaluate the impact of data leakage on model performance.

\textbf{Main findings and implications.} Based on our experimental results, we have the following main findings: (1) \textit{LCMs perform differently on different tasks.} In code generation of Java, CodeLlama-34B achieve the highest CodeBLEU score of 34.08, while in API recommendation of Java, CodeLlama-13B achieves the highest F1 score of 48.31. (2) \textit{Rich context information can greatly enhance the performance of LCMs.} For instance, compared to the basic context, incorporating full contextual information increase the average CodeBLEU scores of all LCMs in Java and Python code generation by 70.73\% and 31.90\%, respectively. (3) \textit{LCMs show inconsistent performance at different timestamps, specifically performing better on data that has been leaked.} For instance, compared to non-leaked data, in Java and Python code generation on leaked data, the average CodeBLEU scores of LCMs increase by 1.22 and 3.10, respectively.

ComplexCodeEval, its construction tools, and all experimental results have been open-sourced \cite{dataset_extract} to aid researchers and developers in better evaluating and optimizing LCMs. To address potential data leakage with the emergence of new models, we plan to regularly update our benchmark every six months, similar to the approach taken by LiveCodeBench \cite{jain2024livecodebench}, to ensure compatibility with the latest mainstream LCMs.

In summary, this paper makes the following contributions: 
\begin{itemize}
\item We introduce ComplexCodeEval, a benchmark suitable for multiple downstream tasks, which can reflect the various programming environment and deliberately avoid data leakage problems.
\item We evaluate the performance of ten popular models on four tasks, revealing the programming capabilities of LCMs more comprehensively.
\item We reveal the impact of context and time on the performance of LCMs in code generation and code completion.
\end{itemize}

\section{Background And Related Work}

\begin{table*}[ht]
\centering
\caption{The comparison between existing benchmarks and ComplexCodeEval. CC, M/A, NL, \wcz{and ADL} indicate cyclomatic complexity, Manual/Automated, nature language and avoiding data leak, respectively.}
\begin{tabular}{cccccccccc}
\hline
 &
  \multicolumn{5}{c}{Code Distribution} &
   &
   &
   &
   \\ \cline{2-6}
\multirow{-2}{*}{Benchmark} &
  \#Repos &
  \#Samples &
  \#LOC &
  \#Tokens &
  \#CC &
  \multirow{-2}{*}{Annotation} &
  \multirow{-2}{*}{M/A} &
  \multirow{-2}{*}{ADL} &
  \multirow{-2}{*}{Task} \\ \hline
Concode &
  - &
  2,000 &
  1.00 &
  30.59 &
  1.49 &
  NL, Code &
  A &
  \ding{55} &
  CG \\
CoNaLA &
  - &
  500 &
  1.11 &
  14.28 &
  0.00 &
  NL, Code &
  A &
  \ding{52} &
  CG \\
methods2test &
  1,017 &
  78,388 &
  - &
  - &
  - &
  Code, Test &
  A &
  \ding{55} &
  TCG \\
APIBench-C &
  3,700 &
  - &
  - &
  - &
  - &
  Code &
  A &
  \ding{55} &
  AR \\
APPS &
  - &
  10,000 &
  16.46 &
  140.61 &
  2.27 &
  NL, Code, Example &
  A &
  \ding{52} &
  CG \\
MBPP &
  - &
  974 &
  6.68 &
  48.60 &
  2.79 &
  NL, Code &
  M &
  \ding{52} &
  CG \\
HumanEval &
  - &
  164 &
  8.95 &
  57.57 &
  3.59 &
  NL, Code, Signature &
  M &
  \ding{52} &
  CG \\
DS-1000 &
  - &
  1,000 &
  3.66 &
  43.34 &
  0.52 &
  NL, Code, Context &
  A &
  \ding{52} &
  CG \\
CrossCodeEval &
  1,002 &
  9,928 &
  - &
  - &
  - &
  NL, Code, Context &
  A &
  \ding{52} &
  CC \\
ClassEval &
  - &
  100 &
  45.70 &
  123.70 &
  - &
  NL, Code, Depend &
  M &
  \ding{52} &
  CG \\
CoderEval-Java &
  10 &
  230 &
  9.17 &
  66.42 &
  3.10 &
   &
   &
   &
   \\
CoderEval-Python &
  43 &
  230 &
  20.64 &
  119.26 &
  4.52 &
  \multirow{-2}{*}{NL, Code, Depend, Context} &
  \multirow{-2}{*}{M} &
  \multirow{-2}{*}{\ding{55}} &
  \multirow{-2}{*}{CG} \\ \hline
\rowcolor[HTML]{DAE8FC} 
ComplexCodeEval-Java &
  1,055 &
  3,897 &
  31.72 &
  251.45 &
  6.52 &
  \cellcolor[HTML]{DAE8FC} &
  \cellcolor[HTML]{DAE8FC} &
  \cellcolor[HTML]{DAE8FC} &
  \cellcolor[HTML]{DAE8FC} \\
\rowcolor[HTML]{DAE8FC} 
ComplexCodeEval-Python &
  2,107 &
  7,184 &
  38.21 &
  293.64 &
  7.57 &
  \multirow{-2}{*}{\cellcolor[HTML]{DAE8FC}\begin{tabular}[c]{@{}c@{}}NL, Code, Depend, Context,\\ Repository, Path, Test, Time\end{tabular}} &
  \multirow{-2}{*}{\cellcolor[HTML]{DAE8FC}A} &
  \multirow{-2}{*}{\cellcolor[HTML]{DAE8FC}\ding{52}} &
  \multirow{-2}{*}{\cellcolor[HTML]{DAE8FC}\begin{tabular}[c]{@{}c@{}}CG, CC,\\ TCG, AR\end{tabular}} \\ \hline
\end{tabular}
\label{tab:dataset_compare}
\end{table*}

\subsection{Large Code Models}

LLMs are widely applied in the coding domain for diverse tasks such as code generation, code completion, test case generation, and API recommendation. A specialized subset of LLMs, known as LCMs, are trained on code corpora and instructions. Notable examples of LCMs, such as DeepSeek-Coder \cite{guo2024deepseek} and Starcoder2 \cite{li2023starcoder}, supporting context windows of up to 16k tokens. These models claim proficiency in project-level code generation and completion tasks, demonstrating strong performance across several established benchmarks. Additionally, numerous LCMs have been proposed, including Codellama \cite{roziere2023code}, WizardCoder \cite{luo2023wizardcoder}, and others \cite{allal2023santacoder,wei2023magicoder,zhong2023codegen}.

\subsection{Benchmarks for Code-Related Tasks}

Existing benchmarks for code-related tasks are often oriented towards different downstream tasks. In code generation tasks, LCMs generate code snippets or entire functions based on given nature language descriptions, streamlining the initial development process. Benchmarks for code generation, such as HumanEval \cite{chen2021evaluating} and MBPP \cite{austin2021program}, evaluate the performance of LCMs in generating Python functions from natural language descriptions and function signatures. However, these benchmarks are limited to standalone functions (i.e., the function has no external dependencies). DS-1000 \cite{lai2023ds} expands the scope to tasks involving third-party data science libraries but is restricted to seven libraries with an average code length of 3.66 lines. CoderEval \cite{yu2024codereval} addresses repository-level code generation for both standalone and non-standalone functions, but its evaluation datasets come from only a small number of code repositories (CoderEval is curated from 53 code repositories in total). 

Apart from code generation, code completion, which aims to predict the next part of a code snippet, is also another popular downstream task that has gained a lot of attention. A recently proposed benchmark for this task is CrossCodeEval \cite{ding2023crosscodeeval}, which is used to evaluate LCMs' capabilities in cross-file code completion. However, in real-world software development, the application typically utilize built-in or third-party libraries to realize certain functions. Hence, when developers are utilizing LCMs to perform code completion, there is a high chance that APIs are involved in the code completion task, which is commonly known as API recommendation. API recommendation aids developers in finding and using the most appropriate APIs for their specific needs. Benchmarks for API recommendation tasks, such as APIbench \cite{peng2021revisiting}, facilitate the assessment of query-based and code-based API recommendation scenarios. Apart from that, another important downstream task is test case generation, which leverages LCMs to automatically create test cases, ensuring that the code functions as expected and helping to maintain software reliability.
%another downstream task that is important is test case generation, which aims to utilize LCMs to automatically create test cases, ensuring that the code functions as expected, ensuring and help maintaining software reliability.
Method2Test \cite{tufano2020unit} is a dataset for test case generation, containing 780,944 instances from 9,410 Java projects.

\autoref{tab:dataset_compare} presents twelve different benchmarks, detailing their size, code scale, code complexity, annotation information, collection methods, and target tasks. For comparison, our constructed benchmark, ComplexCodeEval, is displayed in the last two rows.
% shows twelve different benchmarks, including their size, code scale, code complexity, annotation information, collection method, and target tasks, with our constructed benchmark ComplexCodeEval shown in the last 2 rows for comparison. 
While the aforementioned benchmarks have gained significant popularity among researchers for evaluating the performance of LCMs on specific downstream tasks, they exhibit three major limitations. First, they typically focus on only a few specific code repositories or are based on synthetic datasets created manually, rather than real-world repositories, as seen in HumanEval, MBPP, and CoderEval. This lack of broader repository coverage may result in discrepancies between model performance in practical applications and benchmark results. Second, current benchmarks primarily assess isolated code-related tasks, failing to provide a holistic evaluation of LCMs' capabilities across multiple dimensions. Lastly, several benchmarks, including CoderEval, Method2Test, and APIbench, have not adequately addressed data leakage issues, potentially leading to evaluation biases.
% It can be observed that while all the aforementioned benchmarks have gained widespread popularity among researchers to evaluate the performance of LCMs for specific downstream tasks, there are three major limitations of existing benchmarks. First, they typically cover only a few or not specific code repositories, lacking evaluations on a broader range of repositories, such as HumanEval, MBPP and CoderEval. These limitations may lead to discrepancies between models' performance in practical applications and benchmark test results. Additionally, current evaluation benchmarks primarily concentrate on isolated code-related tasks, failing to provide a comprehensive evaluation of the capabilities of LCMs across multiple dimensions. Furthermore, some benchmarks (such as CoderEval, Method2Test, and APIbench) have not effectively mitigated data leakage issues, which may result in evaluation biases.

To address these limitations, we introduce ComplexCodeEval, a novel benchmark that provides comprehensive evaluations across various downstream coding tasks, including code generation, API recommendation, and test case generation, making it a more robust and versatile tool for assessing the diverse capabilities of LCMs. ComplexCodeEval is derived from a wide range of GitHub repositories to ensure it reflects real-world coding scenarios, and it incorporates timestamps to mitigate data leakage. Compared to existing benchmarks, ComplexCodeEval offers broader coverage and more comprehensive evaluations, providing a better reflection of model performance in complex development environments.

\section{ComplexCodeEval}

\begin{figure} [t]
  \centering
  \includegraphics[width=\linewidth]{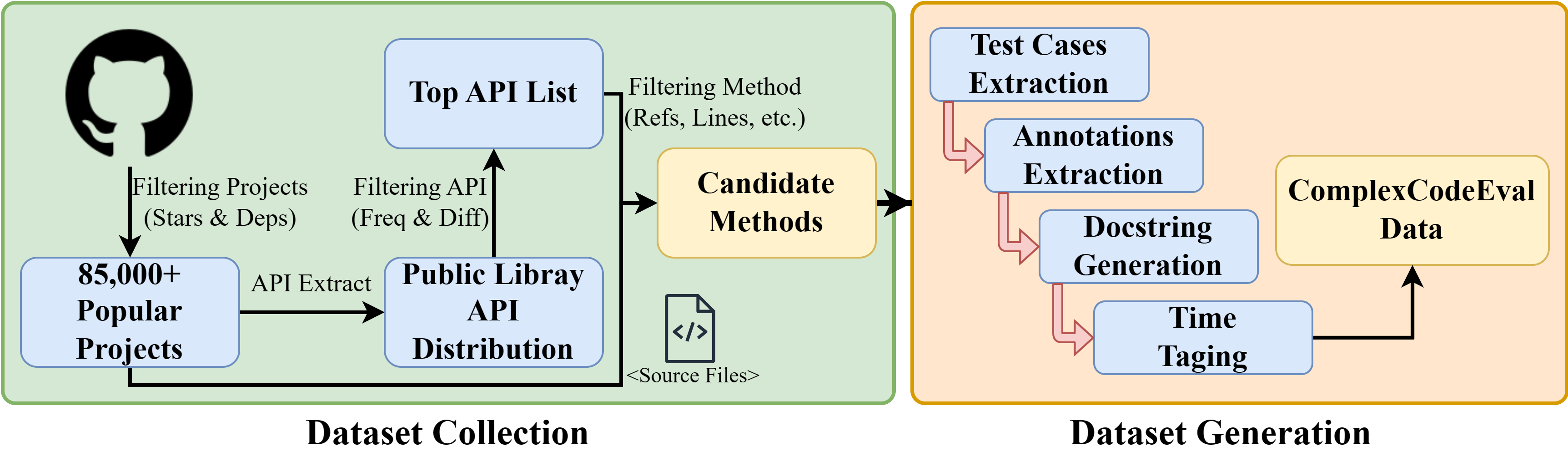}
  \caption{The process of ComplexCodeEval construction.}
  \label{fig:ComplexCodeEval_contruct}
\end{figure}

% \begin{figure*} [ht]
%     \centering
%     \includegraphics[width=\textwidth]{figures/dataset_overview.pdf}
%     \caption{An overview of ComplexCodeEval.}
%     \label{fig:data_overview}
% \end{figure*}
% \begin{figure} []
%     \centering
%     \includesvg[width=\linewidth]{figures/dataset_overview.svg}
%     \caption{An \wcz{example} of ComplexCodeEval.}
%     \label{fig:data_overview}
% \end{figure}

% In this section, we will elaborate on the construction process of ComplexCodeEval, as depicted in \autoref{fig:ComplexCodeEval_contruct}. The construction of ComplexCodeEval is a fully automated pipeline that comprises three primary stages: dataset collection, dataset generation, and evaluation process. In the following subsections, we will describe the details of each stage in detail.

% In this section, we divide the content into three subsections. The first two subsections describe the construction process of ComplexCodeEval (as shown in \autoref{fig:ComplexCodeEval_contruct}), including the data collection and dataset generation processes. The third subsection provides a detailed introduction to the features of ComplexCodeEval.

This section is divided into three subsections. The first two subsections outline the construction process of ComplexCodeEval (as illustrated in \autoref{fig:ComplexCodeEval_contruct}), covering data collection and dataset generation. The third subsection provides a detailed overview of the key features of ComplexCodeEval.

\subsection{Dataset Collection}

\subsubsection{Library Selection.}
% To ensure that ComplexCodeEval closely mirrors real-world development environments and reflects diversity, we integrate popular libraries from Maven (PyPI) based on Libraries.io and actual requirements from our business development processes (derived from survey responses). First, we accessed the Libraries.io API and sorted libraries according to their SourceRank. SourceRank considers multiple factors, including repository activity (e.g., commit frequency, release frequency, issue and pull request handling) and package popularity (e.g., download counts, dependency counts, stars, and forks). For both Python and Java, we selected the top 100 most popular libraries.

To ensure that ComplexCodeEval is close to complicated development environments and reflects diversity, we collect popular libraries from Maven (PyPI) through Libraries.io.
%combine popular libraries from Maven (PyPI) (sourced from Libraries.io) and actual requirements from business development processes (sourced from surveys among developers). 
First, we access the Libraries.io API and sort the libraries based on their SourceRank. SourceRank takes into account multiple factors, including repository activity (e.g., commit frequency, version releases, and issue and pull request handling) and package popularity (e.g., download counts, number of dependencies, stars, and forks). For Python and Java, we select the top 100 most popular libraries respectively. Next, we manually check the collected libraries, excluding tools that could not be integrated through library dependencies (e.g., command-line tools), ultimately filtering down to 69 popular Java frameworks and 55 popular Python packages. These libraries span several broad domains, for instance, web development and network communication (e.g., Java's Spring and Python's Django), data processing and persistence (e.g., Java's MyBatis and Python's pandas), and distributed systems and microservice architectures (e.g., Java's Apache Dubbo and Python's celery).

% Next, we conduct a survey among developers in various departments of an IT company, involving 20 junior to senior-level developers. Due to confidentiality issues, we are unable to disclose the identity of the company and its employees. The survey aims to collect data on the libraries they most frequently use in their daily development. This survey ensure that the selected libraries cover a wide range of development tasks and technology stacks, reflecting industry best practices while also meeting actual development needs.

% Finally, we manually review each library collected in the first two stages, excluding tools that could not be integrated through library dependencies (e.g., command-line tools), ultimately filtering down to 69 popular Java frameworks and 55 popular Python packages. These libraries span several broad domains, including web development and network communication (e.g., Java's Spring and Python's Django), data processing and persistence (e.g., Java's MyBatis and Python's pandas), distributed systems and microservice architectures (e.g., Java's Apache Dubbo and Python's celery), serialization and data exchange (e.g., Java's Jackson and Python's ujson), security and encryption (e.g., Java's Bouncy Castle and Python's cryptography), monitoring and metrics (e.g., Java's Prometheus and Python's prometheus\_client), and logging and utilities (e.g., Java's Log4j and Python's loguru).

\subsubsection{Repository Selection.}
\wcz{We collect Java and Python repositories from GitHub with over 99 stars and obtain more than 85,000 repositories. Then we analyze the dependencies of each repository.} Specifically, we first extract the external libraries that the code repositories depend on by parsing all the configuration files in the repository (e.g., Java's pom.xml; Python's requirements.txt and setup.py). To ensure that the extracted dependencies are more comprehensive and accurate, we also retrieve dependencies from the repositories' corresponding Software Bill of Materials (SBOM) on GitHub, combining both sources to obtain the full dependency information. Based on the dependency information, we retain the repositories that depend on the selected libraries. Ultimately, we retain 9,169 Java projects and 27,178 Python projects.

\subsubsection{Candidate Methods Extraction.}
We first manually define API filtering rules for each selected library, focusing on the common prefixes of the APIs within each library. These rules are implemented by specifying Accept and Reject fields for each library. The Accept field lists common prefixes for APIs belonging to the target library, while the Reject field excludes APIs that share similar prefixes but belong to different libraries. Next, for each selected project, we construct an abstract syntax tree (AST) for every code file. By traversing the ASTs and applying the filtering rules, we determine which APIs from the selected libraries are called in each code file and the corresponding frequencies. We then aggregate this information to obtain the API usage frequencies across all projects, selecting the top 100 (if available) most frequently called APIs for each library. For the projects that call these APIs, we construct ASTs for their code files and model each project, retaining all test functions, as well as non-test functions that have more than 10 lines of code.

\subsection{Dataset Generation}
\subsubsection{Test Cases Extraction.}

\begin{figure} [t]
    \centering
    \includegraphics[width=\linewidth]{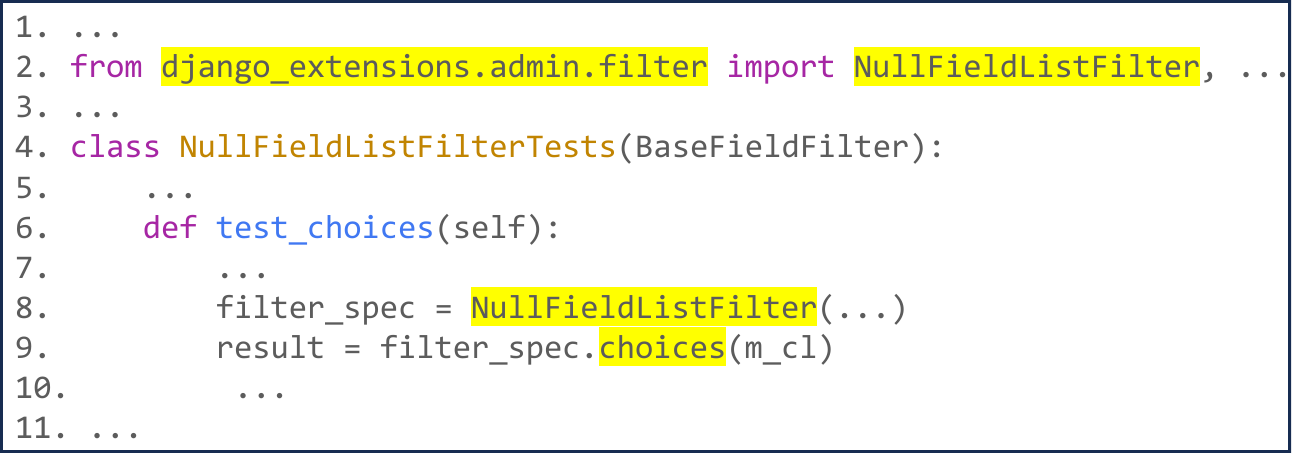}
    \caption{An example of a Python test function. The \textit{choices} function is invoked by \textit{filter\_spec} (line 9), where \textit{filter\_spec} is an instance of \textit{NullFieldListFilter}. \yun{Considering that} \textit{NullFieldListFilter} is imported \yun{from the \textit{django\_extensions.admin.filter.NullFieldListFilter} class} (line 2),
    % . Therefore, 
    the original path of \textit{choices} is \yun{set as the imported class.}}
    \label{fig:sample_test_function}
\end{figure}
We utilize the project modeling obtained from the \textit{Candidate Methods Extraction} stage to match functions with their corresponding test functions. Specifically, for Java projects, we match them through the following steps: (1) Traverse each non-test file in the project and match it with its corresponding test file by filename. (2) Within each matched non-test file, traverse each function entity and match it with the corresponding test function by function name. (3) In each matched test function, traverse all its function calls and determine whether it calls the corresponding non-test function by function name and parameters. Through these steps, we ultimately obtain the pairs of functions and test functions.

Since Python test functions do not have a uniform standard, and their parameter types and quantities are indeterminate, the same matching method used for Java projects cannot be directly applied. Therefore, we take the following steps for Python projects: (1) Match non-test files and test files by filename. (2) In the matched test files, traverse each test function. (3) For each test function, reconstruct the original path of each function call (e.g.,  as shown in \autoref{fig:sample_test_function}, the original path of the \textit{choices} function is 
\textit{django\_extensions.ad\allowbreak min.filter.NullFieldListFilter.choices}). (4) Traverse each function in the corresponding non-test file, match the function path with the original path. Through these steps, we obtain the pairs of functions and test functions.

\subsubsection{Annotations Extraction.}

\begin{figure} [t]
    \centering
    \includegraphics[width=\linewidth]{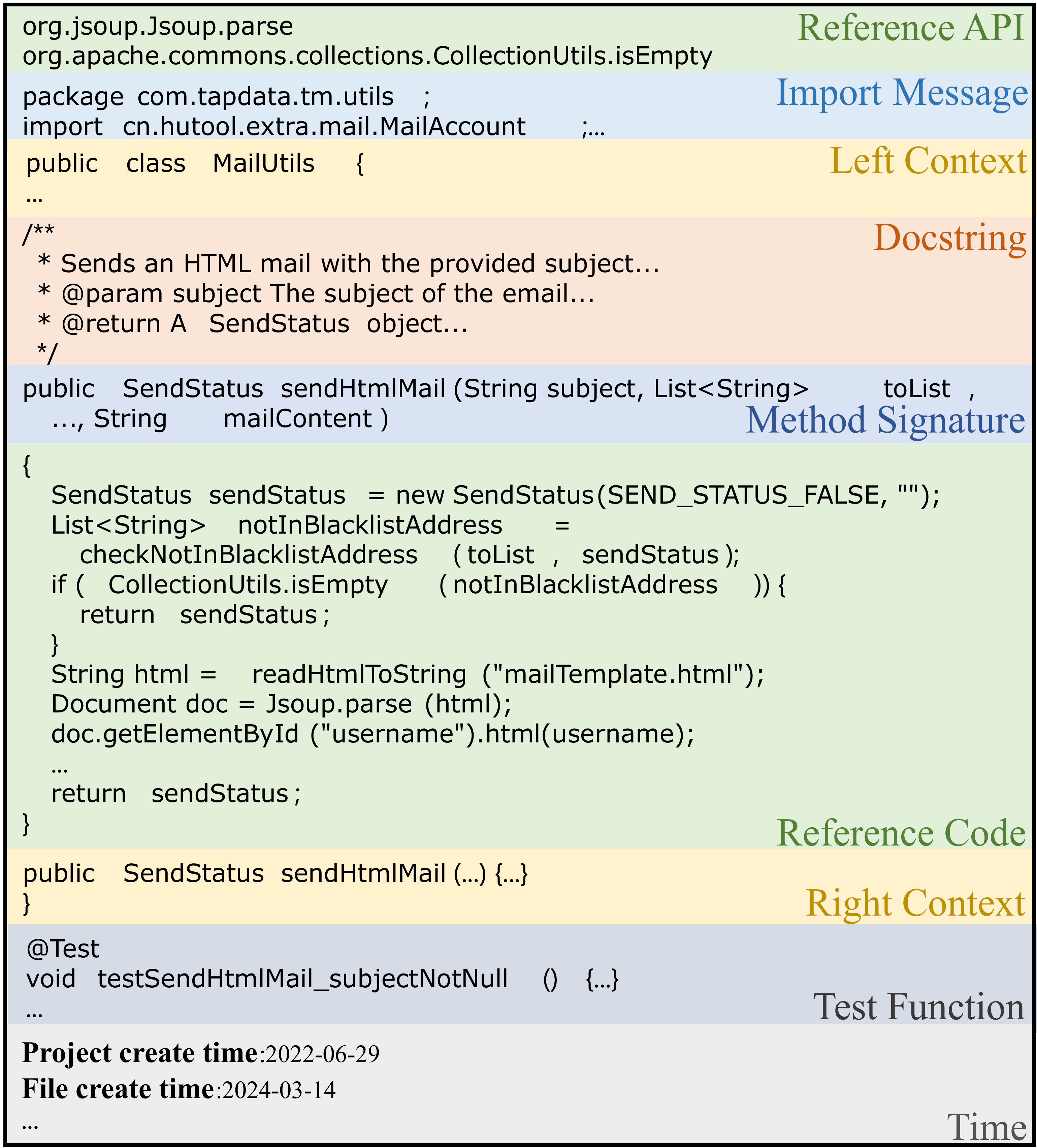}
    \caption{An \wcz{example} of ComplexCodeEval.}
    \label{fig:data_overview}
\end{figure}

To ensure that ComplexCodeEval is suitable for multiple downstream tasks, we make considerable efforts to extract metadata related to different downstream tasks. Specifically, we use the functions extracted from the \textit{Test Cases Extraction} stage as function samples and further analyze the corresponding repositories for each function, annotating each sample with multiple attributes, such as: % (1) original repository information (e.g., repository owner, repository name, and version), (2) function namespace (e.g., file path of the function, module, and class where the function resides), (3) function information (e.g., function name, parameters, signature), (4) file context, (5) original documentation comments (i.e., natural language description of the function), (6) function implementation, (7) function dependencies (i.e., external dependencies called within the function implementation), (8) corresponding test functions, (9) temporal information (e.g., project creation and update times).
reference API (i.e., the popular APIs called by the samples), import information, context, docstring (i.e., natural language description of the function), function signature, code implementation, test function. An example of a sample can be seen in \autoref{fig:data_overview}.

To ensure the diversity of ComplexCodeEval samples, we ensure that each sample under an API comes from different projects (i.e., the same API can only extract one sample in a project). When the similarity between samples under one API and another API exceeds half of its own sample count, we randomly remove one of the APIs and its corresponding samples. Furthermore, we reselect high-frequency APIs within the corresponding frameworks based on API call frequency for sample extraction. Finally, we perform deduplication of function implementations for the samples using the exact matching method.

\subsubsection{Docstring Generation.}
% \lyq{When applying LCMs to code-related tasks, especially code generation, the quality of the prompt significantly affects the quality of the generated code. If the prompt does not accurately describe the target code, the quality of the code generated by the LCMs will not be high.} To investigate the quality of the original documentation comments, we conducted a sample inspection and found that the quality of these comments varied greatly. Many documentation comments were relatively short, containing only a simple one-sentence description of the target function. Their expressions were often unclear and incomplete.

% \lyq{Due to the time-consuming and labor-intensive nature of manually writing documentation comments, and inspired by the strong capability of LCMs in generating code annotations \cite{geng2024large}, we decided to use LCMs to generate documentation comments. Specifically, }we designed an initial prompt to have the LCMs generate documentation comments. We then manually evaluated the quality of these comments and optimized the prompt based on the feedback. This process was repeated iteratively, continually refining the prompt until the generated documentation comments met the desired quality standards. Ultimately, we obtained a set of high-quality, accurate, and comprehensive documentation comments for the sample functions. However, to maintain the authenticity of the samples, we retained the original documentation comments and added the LCMs-generated comments for each sample.

\begin{figure} [t]
    \centering
    \includegraphics[width=\linewidth]{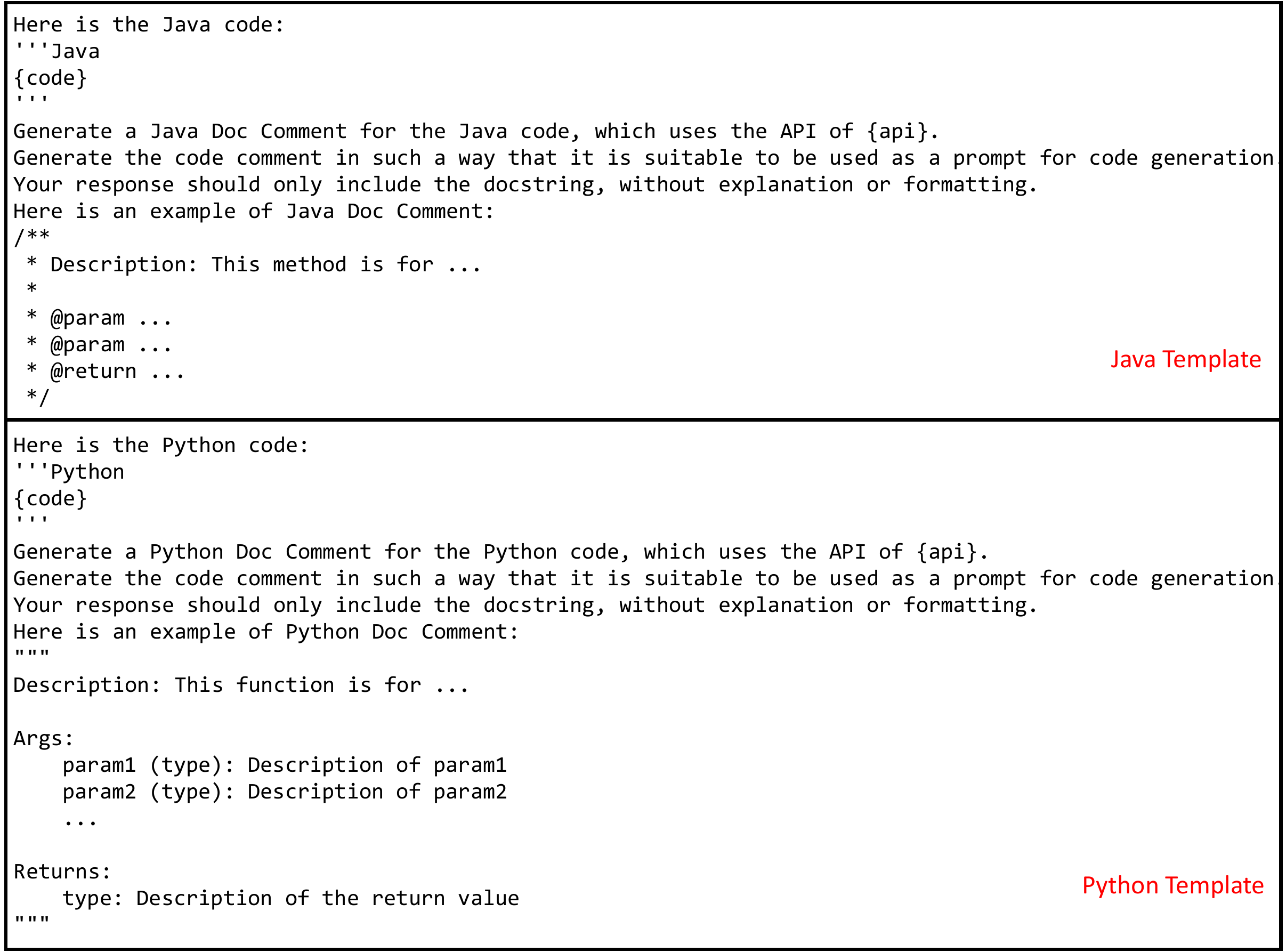}
    \caption{The prompt template used for generating docstrings.}
    \label{fig:prompt_template}
\end{figure}

\begin{figure} [t]
    \centering
    \includegraphics[width=\linewidth]{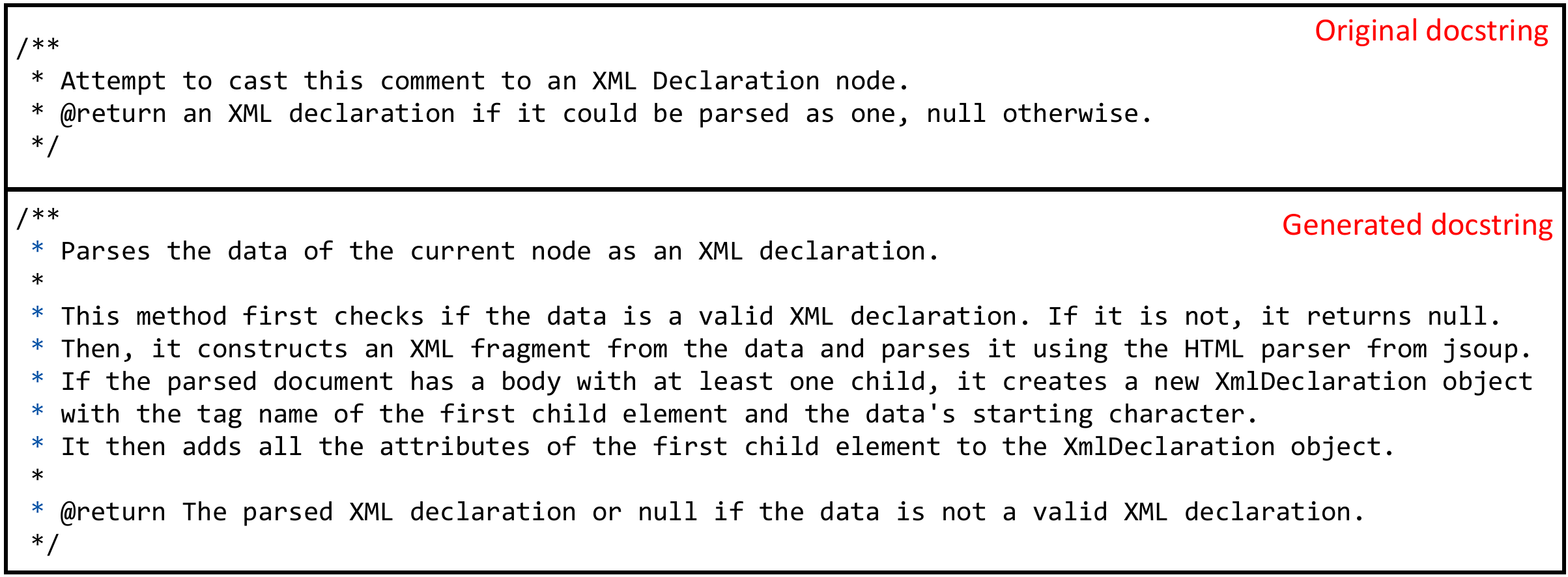}
    \caption{An example of the original docstring and the LLM-generated docstring.}
    \label{fig:docstring}
\end{figure}

Due to the impact of prompt quality on the code generated by LCMs, it is essential to ensure that the docstrings provided in ComplexCodeEval comprehensively and accurately describe the functions (where docstring is usually used as the prompt in benchmarks). Therefore, we investigate the quality of the original docstrings. Specifically, we sample and check the original docstrings (an example can be seen in \autoref{fig:docstring}) and find that their quality varied widely. Many docstrings are relatively short, containing only a brief description of the target function, and are often ambiguous and incomplete.

Manually writing docstring is both time-consuming and labor-intensive. Inspired by the strong capabilities of LLMs in generating code comments \cite{geng2024large}, we use a LCM (DeepSeek-Coder-33B) to generate docstrinsg for all of the candidate questions in ComplexCodeEval (see Section \ref{sec:docstring_compare} for further analysis). Specifically, we design an initial prompt template to instruct the model to generate docstrings, and we manually evaluate the effectiveness of the prompt template through an iterative process. To ensure its robustness, we randomly sample 100 examples, assessing the generated docstrings based on three criteria: (1) The docstrings should only include necessary descriptions without extraneous content, (2) the parameters and return values must accurately match those of the original function, and (3) the docstring should clearly and correctly reflect the function’s purpose and content, as verified independently by the first and second authors. We repeat this process, refining the prompt template until it consistently produces high-quality docstrings. The final prompt template can be seen in \autoref{fig:prompt_template}. 
% Specifically, a high-quality docstring must explain the purpose of the code, clearly describe the input parameters and their types, outline the output and its type, and provide any necessary context or additional information that enhances the understanding of the function.
Additionally, to maintain the authenticity of the samples, we retain the original docstring and add the LLM-generated docstring (an example of docstring can be seen in \autoref{fig:docstring}) for each sample.

\subsubsection{Time tagging.}

% \begin{figure*} [t]
%   \centering
%   \includegraphics[width=\textwidth]{figures/sample_distribution.pdf}
%   \caption{Sample distribution of project creation time, project update time, file creation time, file update time, and function update time.}
%   \label{fig:sample_distribution}
% \end{figure*}

% \lyq{In the evaluation of LLMs, the issue of data leakage has always been a significant risk and challenge. Previous works have discussed how to avoid data leakage. For instance, CoderEval \cite{yu2024codereval} uses manually annotated docstrings to avoid data leakage, while EvoCodeBench \cite{li2024evocodebench} and LiveCodeBench \cite{jain2024livecodebench} opt for periodic releases of new versions to mitigate this issue.} We have adopted a more flexible approach by adding temporal information to each sample to dynamically address the problem of data leakage.

Data leakage is one of the most crucial problems to address when it comes to the LLM/LCM benchmarks. It is imperative to ensure that the questions in the benchmark are not seen by the model. We adopt a flexible approach by adding timestamps to each function sample to dynamically address the problem of data leakage. Initially, we consider using the repository creation time as the time information for each function sample. However, upon analyzing the distribution of function samples based on repository creation and update times, we find that using only repository time information did not adequately reflect the function sample distribution. Therefore, we obtain more detailed time information by analyzing the git commit history of each sample's corresponding GitHub repository. Specifically, we use the earliest commit time of the file corresponding to the sample as the file creation time, %the most recent commit time on the current branch as the file update time, 
the most recent commit time of the file as the file update time and the most recent commit time involving additions or modifications to the function's code as the function update time (deletions of function code were not considered updates). By using the timestamp information, we can purposefully avoid data leakage issues by considering the model's knowledge cut-off date.

\subsection{Benchmark Characteristic}
Through the aforementioned steps, we construct ComplexCodeEval. Its detailed characteristics are as follows:

\textbf{Flexibility.}
ComplexCodeEval contains rich information, including context, reference APIs, and test functions (as shown in \autoref{fig:data_overview}). These elements can be flexibly combined, enabling ComplexCodeEval to be used for various code tasks such as code generation, code completion, test case generation, and API recommendation. Additionally, various supplementary information can be incorporated into each task to explore the performance of LCMs with different data inputs. Moreover, ComplexCodeEval provides multiple timestamps, including the creation and update times at the project, file, and method levels. We can leverage this information to adjust the data to meet our needs, such as avoiding data leakage issues.

\textbf{Scale.}
\autoref{tab:dataset_compare} presents the overall data scale information of ComplexCodeEval for comparison with other benchmarks. The results show that ComplexCodeEval comprises 3,897 samples from 1,055 Java projects and 7,184 samples from 2,107 Python projects. The average lines of code in ComplexCodeEval-Python reaches 38.21, second only to ClassEval's 45.70 (notice that ClassEval is at the class level, containing multiple functions, whereas ComplexCodeEval is at the function level, containing only a single function). Furthermore, the average cyclomatic complexity of the Python and Java samples in ComplexCodeEval are 7.57 and 6.52, respectively, higher than all existing datasets. These results indicate the challenge of tasks involved in our benchmark.

\section{Experimental setup}

With ComplexCodeEval, our experiment aims to answer the following research questions:
\begin{itemize}
\item {\textbf{RQ1}}: How do LCMs perform on ComplexCodeEval varies across different tasks?
\item {\textbf{RQ2}}: How does different contextual information impact LCMs performance on these tasks?
\item {\textbf{RQ3}}: How does data leakage affect the effectiveness of LCMs on these tasks?
\end{itemize}

\subsection{Model Selection}
% \usepackage{booktabs}
\begin{table}[t]
\centering
\caption{\yun{Selected} LCMs.
% seletion.
}
\begin{tabular}{@{}lcccc@{}}
\toprule
Model          & size & \multicolumn{1}{c}{time} & \multicolumn{1}{c}{instruct} & \multicolumn{1}{c}{base}   \\ \midrule
DeepSeek Coder & 33B  & 2023-01-01               & \ding{51}   & \ding{51} \\
DeepSeek Coder & 6.7B & 2023-01-01               & \ding{51}   & \ding{51} \\
DeepSeek Coder & 1.3B & 2023-01-01               & \ding{51}   & \ding{51} \\
StarCoder 2    & 15B  & 2023-09-14               & \ding{51}   & \ding{51} \\
StarCoder 2    & 7B   & 2023-09-14               & \ding{55}   & \ding{51} \\
StarCoder 2    & 3B   & 2023-09-14               & \ding{55}   & \ding{51} \\
CodeLLaMa      & 34B  & 2023-01-01               & \ding{51}   & \ding{51} \\
CodeLLaMa      & 13B  & 2023-01-01               & \ding{51}   & \ding{51} \\
CodeLLaMa      & 7B   & 2023-01-01               & \ding{51}   & \ding{51} \\
GPT-3.5-Turbo  & N/A  & 2021-10-01               & \ding{51}   & \ding{55} \\ \bottomrule
\end{tabular}
\label{tab:LCMs}
\end{table}
As shown in \autoref{tab:LCMs}, we select ten distinct LCMs to assess their performance on code intelligence tasks. These models include four categories: the StarCoder2 family (SC2), the Codellama family (CL), the DeepSeek-Coder family (DSC), and one closed-source model (GPT-3.5-Turbo). These models encompass various scales and training objectives, ranging from small-scale to large-scale models, covering state-of-the-art large code model technologies. The selection of models is based on their performance in existing research and their availability.

\subsection{Task Design}

To comprehensively assess the performance of LCMs, we identify four representative tasks: code generation, code completion, test case generation, and API recommendation. Below, we describe each of these tasks in detail. %The input-output format for each task detailed in Appendix C.

\textbf{Code Generation} is a common task to evaluate the capabilities of LCMs to translate natural language into code. In our experiment, the model is furnished with a function signature and its corresponding docstring, and is required to generate a correct implementation. This task simulates a scenario where developers write documentation describing the functionality and signature of a function, using LCMs to automatically generate the function's implementation code to enhance development efficiency.

\textbf{Code Completion} is employed to assess the code completion capabilities of LCMs, which showcases how LCMs can be used to predict and complete code based on an initial portion provided by developers, improving programming efficiency and code quality. Inspired by previous research \cite{nguyen2019focus}, we provide the initial half of the function to the LCMs, and the LCMs need to complete the remaining portion.

\textbf{Test Case Generation} aims to evaluate the LCMs' comprehension of the original function's functionality. In this task, the model is provided with the original function and the function signature of the test cases. The LCM is subsequently required to generate the specified test case to assess specific functionality or segments of the code. This scenario illustrates how LCMs can automatically create test cases that validate code functionalities, reducing manual effort and enhancing test coverage.

\textbf{API Recommendation} tasks are generally categorized into query-based API recommendation and code-based API recommendation. This paper focuses on the latter, where the model is provided with a code snippet and is required to recommend an appropriate public library API. This scenario highlights how LCMs can help developers quickly find suitable APIs based on the provided code snippet, enhancing development efficiency and reducing the time spent searching and selecting APIs.
\subsection{Annotation selection}
To accommodate the requirements of the four selected tasks, we select the following annotations from ComplexCodeEval, accompanied by their corresponding explanations.
\begin{itemize}
\item \textbf{Reference API (A)} includes the public library APIs invoked by the target function. It serves as a reference during the evaluation phase of API recommendation tasks.
\item \textbf{Import Information (I)} encompasses the libraries imported within the file and the package information of the file itself. It can be provided as supplementary information to LCMs for inference across various tasks.
\item \textbf{File Context (C)} consists of all code snippets within the file that are outside the target function. It can be used as additional information to aid LCMs in reasoning.
\item \textbf{Docstring (D)} refers to the documentation comments of the target function, typically describing the function's logic, parameters, and return values. It is used as part of the prompt in code generation tasks.
\item \textbf{Method Signature (S)} is the function signature of the target function, also included as part of the prompt in code generation tasks.
\item \textbf{Reference Code (R)} is the reference implementation of the target function from the original project. It is used as a ground truth in the evaluation phase of code generation tasks and can also serve as a prompt for code completion, test case generation, and API recommendation tasks.
\item \textbf{Code Snippet (CS)} refers to the lines of code preceding the target completion line, which are used for code completion.
\item \textbf{Test Function (T)} is the corresponding test function for the target function from the original project. It is used as a reference during the evaluation phase of test case generation tasks.
\item \textbf{Dependency (Dep)} includes all external dependency functions within the target function. It can be provided as additional information for LCMs to reason across various tasks.
\item \textbf{Timestamp (TS)} contains timestamps such as project creation and file creation dates, which can be utilized for dataset partitioning.
\end{itemize}

\subsection{Evaluation Metrics}
For the first three tasks (code generation, code completion, test case generation), we use CodeBLEU \cite{ren2020codebleu} (or BLEU \cite{association1996proceedings} in code completion) and ES (Edit Similarity) as the evaluation metric. CodeBLEU is specifically designed to evaluate the quality of code generation, considering multiple aspects like BLEU, AST (abstract syntax tree), data flow, and naming entities, providing a comprehensive reflection of generated code quality. The computation of CodeBLEU score includes the following steps:
\begin{itemize}
\item {\textbf{BLEU}}: Evaluates n-gram and weighted n-gram overlap between generated code and reference code.
\item {\textbf{Syntax Match}}: Compares the abstract syntax tree structures of the generated and reference codes.
\item {\textbf{Data Flow Match}}: Analyzes data flow similarity between generated and reference codes.
\item {\textbf{Average Score}}: Calculates the average value of the aforementioned scores.
\end{itemize}

For the API recommendation task, we use Recall and F1 to evaluate the quality of the APIs recommended by the model.

\subsection{Implementation Details}

Our experiments are run on a server with two A100-80G GPUs. To eliminate the effects of random sampling, we adopted a greedy decoding strategy. Specifically, we set the temperature to 0, top\_p to 1, and max\_tokens to 1000 (with the API-recommended setting being 10). For each task, we randomly select 84 to 100 pieces of samples for both Java and Python from the dataset, specifically from data after September 15, 2023, based on the model's knowledge cut-off time (as shown in \autoref{tab:LCMs}). %Compared to existing benchmarks, ComplexCodeEval offers broader coverage and more comprehensive evaluations, better reflecting models' performance in real-world development scenarios.\textbf{[???]}

\begin{table*}[ht]
\centering
\caption{Performance of various LCMs on ComplexCodeEval. Bold numbers on a gray background indicate the maximum values. }
\scalebox{0.9}{
\begin{tabular}{@{}lccccccccccc|c@{}}
\toprule
Metrics  & \multicolumn{1}{l|}{Language} & DSC-33B & DSC-6.7B & DSC-1.3B & SC2-15B & SC2-7B & SC2-3B & CL-34B & CL-13B & CL-7B & GPT-3.5 & $\sigma$ \\ \midrule
\multicolumn{11}{c}{Code Generation}                                                                                                                                                                         \\ \midrule
CodeBLEU & \multicolumn{1}{l|}{Java}   & 33.96                       & 32.23                        & 26.91                        & 32.59                       & 30.83  & 22.46  & \cellcolor{gray!30}{\textbf{34.08}}  & 28.98  & 29.58 & 19.92 & 4.79\\
ES       & \multicolumn{1}{l|}{Java}   & 35.64                       & 34.24                        & 29.70                        & 35.90                       & 31.55  & 26.38  & \cellcolor{gray!30}{\textbf{36.61}}  & 34.04  & 33.61 & 21.66 & 4.78\\
CodeBLEU & \multicolumn{1}{l|}{Python} & 26.29                       & 25.72                        & 21.90                        & 26.70                       & 27.19  & 24.68  & \cellcolor{gray!30}{\textbf{27.54}}  & 19.95  & 25.90 & 26.09 & 2.43\\
ES       & \multicolumn{1}{l|}{Python} & 30.45                       & 29.70                        & 26.00                        & 32.15                       & 32.13  & 30.17  & \cellcolor{gray!30}{\textbf{32.93}}  & 26.92  & 31.09 & 31.05 & 2.24\\ \midrule
\multicolumn{11}{c}{Code Completion}                                                                                                                                                                         \\ \midrule
BLEU & \multicolumn{1}{l|}{Java} & \cellcolor{gray!30}{\textbf{31.86}} & 30.81 & 20.53 & 21.67 & 16.82 & 15.91 & 22.36 & 21.28 & 26.09 & 20.31 & 5.33\\
ES & \multicolumn{1}{l|}{Java} & 33.09 & \cellcolor{gray!30}{\textbf{34.17}} & 26.38 & 24.22 & 17.35 & 20.54 & 26.78 & 25.83 & 29.34  & 24.21 & 5.16\\
BLEU & \multicolumn{1}{l|}{Python} & 16.70 & 15.33 & 10.69 & 12.18 & 11.32 & 11.50 & 13.79 & 10.97 & 11.97 & \cellcolor{gray!30}{\textbf{19.62}} & 2.95\\
ES & \multicolumn{1}{l|}{Python} & \cellcolor{gray!30}{\textbf{29.18}} & 28.45 & 21.77 & 22.89 & 19.47 & 19.91 & 25.48 & 22.79 & 22.99  & 19.61 & 3.47\\ \midrule
\multicolumn{11}{c}{Test Case   Generation}                                                                                                                                                                  \\ \midrule
CodeBLEU & \multicolumn{1}{l|}{Java}   & 29.33                       & 27.05                        & 19.27                        & 25.27                       & 2.57   & 13.10  & \cellcolor{gray!30}{\textbf{29.90}}  & 25.67  & 23.36 & 25.58 & 8.46\\
ES       & \multicolumn{1}{l|}{Java}   & 29.13                       & 29.10                        & 22.99                        & 26.04                       & 5.24   & 15.96  & 29.39  & \cellcolor{gray!30}{\textbf{29.64}}  & 26.24 & 28.54 & 7.89\\
CodeBLEU & \multicolumn{1}{l|}{Python} & 19.40                       & \cellcolor{gray!30}{\textbf{22.87}}                        & 9.79                         & 18.27                       & 14.32  & 13.53  & 21.40  & 20.80  & 19.41 & 17.67 & 4.05\\
ES       & \multicolumn{1}{l|}{Python} & 24.95                       & \cellcolor{gray!30}{\textbf{28.65}}                        & 13.38                        & 25.92                       & 19.74  & 18.37  & 28.00  & 27.70  & 24.80 & 25.54 & 4.95\\ \midrule
\multicolumn{11}{c}{API Recommendation}                                                                                                                                                                      \\ \midrule
F1       & \multicolumn{1}{l|}{Java}   & 44.57                       & 41.85                        & 34.00                        & 32.20                       & 42.15  & 39.43  & 38.92  & \cellcolor{gray!30}{\textbf{48.31}}  & 45.75 & 47.21 & 5.38\\
Recall   & \multicolumn{1}{l|}{Java}   & 44.72                       & 42.00                        & 33.71                        & 32.29                       & 42.18  & 39.57  & 38.73  & \cellcolor{gray!30}{\textbf{47.56}}  & 45.75 & 46.67 & 5.26\\
F1       & \multicolumn{1}{l|}{Python} & 49.76                       & 43.18                        & 37.47                        & 47.48                       & 45.61  & 41.98  & \cellcolor{gray!30}{\textbf{52.24}}  & 42.20  & 36.09 & 43.88 & 5.04\\
Recall   & \multicolumn{1}{l|}{Python} & 49.84                       & 42.97                        & 37.35                        & 48.03                       & 46.31  & 42.29  & \cellcolor{gray!30}{\textbf{52.14}}  & 42.83  & 36.36 & 44.16 & 5.05\\ \bottomrule
\end{tabular}}
\label{tab:rq1}
\end{table*}

\section{Evaluation Result}

% In this section, we provide a detailed overview of our experimental results and findings on ten popular LCMs. %Note that in our table, the maximum values (i.e., the best-performing among the various LCMs) or the difference between the two data points will be displayed in bold. Red indicates the growth relative to the baseline.

\subsection{RQ1: Performance of LCMs on ComplexCodeEval} \label{sec:RQ1}
In \autoref{tab:rq1}, we present the performance of ten LCMs on four different tasks in Python and Java programming languages. From \autoref{tab:rq1}, we can derive the following observations:
\textbf{LCMs still exhibit certain limitations in intricate development scenarios.}
Experimental results indicate that LCMs generally exhibit suboptimal performance across all four tasks. For instance, in the code generation task, the best-performing model, Codellama-34B, achieves CodeBLEU scores of only 27.54 for Python and 34.08 for Java. These scores are benchmarked against the original developer-written code in the collected projects. The low CodeBLEU scores highlight a significant disparity between the code generated by LCMs and the original code. Similarly, in the code completion task, the highest BLEU scores achieved are 19.62 for Python and 31.86 for Java. In the test case generation task, the top CodeBLEU scores are 22.87 and 29.90, respectively. For the API recommendation task, the highest F1 scores are 52.24 for Python and 48.31 for Java. These results indicate that LCMs are still far from being viable for real-world development tasks and suggest the need for supplementary techniques to enhance reasoning capabilities and further advancements in model intelligence.

% It is observed during the experiments that LCMs generally performed poorly across the four tasks. For example, in the code generation task, the best-performing model, Codellama-34B, only achieves CodeBLEU scores of 27.54 and 34.08 in Python and Java, respectively. The benchmark for these scores is the original code written by developers in the collected projects. The low CodeBLEU scores indicate a pronounced gap between the code generated by LCMs and the original code. Similarly, in the code completion task, the highest BLEU scores for Python and Java are only 19.62 and 31.86, respectively. In the test case generation task, the highest CodeBLEU scores are 22.87 and 29.90, respectively. For the API recommendation task, the highest F1 scores are 52.24 and 48.31 for Python and Java, respectively. The poor performance of LCMs on our four real-world development tasks suggests that LCMs are still quite far from being applicable in actual development. It is still necessary to use other techniques to assist LCMs in reasoning and to further develop more intelligent models.

\textbf{Each model exhibits unique capabilities depending on the specific tasks and programming languages.}
We find that each model excels in different tasks during the evaluation process. For instance, although Codellama-34B performs the best in the code generation task for both Python and Java, the best-performing model in code completion is DeepSeek-Coder-33B. In test case generation, Codellama-34B achieves the highest score in Java, while DeepSeek-Coder-6.7B outperforms all other models in Python with a CodeBLEU score of 22.87. For the API recommendation task, Codellama-34B achieves the highest score of 52.24 on the Python dataset but performs poorly on the Java dataset with an F1 score of only 38.92, far inferior to Codellama-13B. Although larger models generally perform better on tasks, these differences indicate that it is difficult to generalize which model is the best across different tasks and languages. A comprehensive evaluation of the models' capabilities in various aspects is necessary, and in practical applications, the most suitable model should be selected based on specific needs.
% \vspace{1em}

% \noindent\fbox{
% \begin{minipage}{\dimexpr\linewidth-2\fboxsep-2\fboxrule}  
% % \setlength{\parindent}{1em}
% \strut
% In summary, by comparing the performance of LCMs on the four public library level tasks, we find that public library level code tasks remain quite challenging for existing models. Furthermore, each model excels in different tasks, and our ComplexCodeEval dataset allows for a more comprehensive assessment of the diverse capabilities of different models.
% \strut
% \end{minipage}  
% }

\finding{1}{Current models continue to display limitations in intricate development scenarios. Furthermore, the performance of different models varies across programming languages and tasks, highlighting the need for a thorough evaluation of each model's capabilities across multiple dimensions.}

\subsection{RQ2: Impact of Contextual Information on LCM's Performance} \label{section:RQ2}
\begin{table*}[t]
\centering
\caption{Performance of LCMs in code generation across different contexts. Bold numbers on a gray background indicate the maximum values and red indicates the growth relative to the baseline.}
\scalebox{0.9}{
\begin{tabular}{@{}lcccccccccc|l@{}}
\toprule
Annotations & \multicolumn{1}{c|}{Language}                    & DSC-33B & DSC-6.7B & DSC-1.3B & SC2-15B & SC2-7B & SC2-3B & CL-34B & CL-13B & CL-7B & Average \\ \midrule
\multicolumn{12}{c}{CodeBLEU}                                                                                                             \\ \midrule
D+S         & \multicolumn{1}{c|}{Python} & 26.29   & 25.72    & 21.90    & 26.70   & 27.19  & 24.68  & \cellcolor{gray!30}{\textbf{27.54}}  & 19.95  & 25.90 & 25.10   \\
D+S+I       & \multicolumn{1}{c|}{Python} & \cellcolor{gray!30}{\textbf{31.13}}   & 29.32    & 22.83    & 29.76   & 28.19  & 27.83  & 31.04  & 25.04  & 28.91 & \makecell{28.23 \textcolor{red}{(\(\uparrow\)12.48\%)}}   \\
D+S+C       & \multicolumn{1}{c|}{Python} & 34.80   & 32.01    & 26.02    & \cellcolor{gray!30}{\textbf{35.34}}   & 28.77  & 30.63  & 31.52  & 26.49  & 29.72 & \makecell{30.59 \textcolor{red}{(\(\uparrow\)21.88\%)}}   \\
D+S+C+Dep   & \multicolumn{1}{c|}{Python} & \cellcolor{gray!30}{\textbf{38.52}}   & 35.68    & 28.52    & 35.87   & 32.18  & 33.17  & 35.15  & 26.40  & 32.44 & \makecell{33.10 \textcolor{red}{(\(\uparrow\)31.90\%)}}   \\ \midrule
D+S         & \multicolumn{1}{c|}{Java}   & 33.96   & 32.23    & 26.91    & 32.59   & 22.01  & 22.46  & \cellcolor{gray!30}{\textbf{34.08}}  & 28.98  & 29.58 & 29.20   \\
D+S+I       & \multicolumn{1}{c|}{Java}   & \cellcolor{gray!30}{\textbf{42.52}}   & 38.89    & 31.42    & 38.36   & 24.37  & 26.02  & 38.80  & 36.87  & 35.04 & \makecell{34.70 \textcolor{red}{(\(\uparrow\)18.83\%)}}   \\
D+S+C       & \multicolumn{1}{c|}{Java}   & 46.90   & \cellcolor{gray!30}{\textbf{47.21}}    & 36.94    & 46.39   & 30.83  & 35.72  & 46.07  & 43.90  & 36.45 & \makecell{41.16 \textcolor{red}{(\(\uparrow\)40.95\%)}}   \\
D+S+C+Dep   & \multicolumn{1}{c|}{Java}   & \cellcolor{gray!30}{\textbf{56.69}}   & 56.33    & 42.96    & 53.76   & 43.64  & 46.68  & 55.35  & 43.75  & 49.53 & \makecell{49.85 \textcolor{red}{(\(\uparrow\)70.73\%)}}   \\ \midrule
\multicolumn{12}{c}{Edit Similarity}                                                                                                      \\ \midrule
D+S         & \multicolumn{1}{c|}{Python} & 30.45   & 29.70    & 26.00    & 32.15   & 32.13  & 30.17  & \cellcolor{gray!30}{\textbf{32.93}}  & 26.92  & 31.09 & 30.17   \\
D+S+I       & \multicolumn{1}{c|}{Python} & 35.57   & 32.30    & 26.20    & 33.99   & 30.97  & 32.11  & \cellcolor{gray!30}{\textbf{36.31}}  & 31.16  & 34.83 & \makecell{32.60 \textcolor{red}{(\(\uparrow\)8.04\%)}}   \\
D+S+C       & \multicolumn{1}{c|}{Python} & \cellcolor{gray!30}{\textbf{39.64}}   & 36.54    & 29.95    & 39.39   & 33.08  & 33.95  & 35.94  & 31.98  & 34.10 & \makecell{34.95 \textcolor{red}{(\(\uparrow\)15.83\%)}}   \\
D+S+C+Dep   & \multicolumn{1}{c|}{Python} & \cellcolor{gray!30}{\textbf{42.06}}   & 39.42    & 32.82    & 40.29   & 35.85  & 36.93  & 39.08  & 32.39  & 35.84 & \makecell{37.19 \textcolor{red}{(\(\uparrow\)23.25\%)}}   \\ \midrule
D+S         & \multicolumn{1}{c|}{Java}   & 35.64   & 34.24    & 29.70    & 35.90   & 24.86  & 26.38  & \cellcolor{gray!30}{\textbf{36.61}}  & 34.04  & 33.61 & 32.33   \\
D+S+I       & \multicolumn{1}{c|}{Java}   & \cellcolor{gray!30}{\textbf{41.73}}   & 37.96    & 31.37    & 36.42   & 24.72  & 27.32  & 39.17  & 39.04  & 34.53 & \makecell{34.70 \textcolor{red}{(\(\uparrow\)7.34\%)}}   \\
D+S+C       & \multicolumn{1}{c|}{Java}   & \cellcolor{gray!30}{\textbf{45.41}}   & 43.95    & 36.04    & 44.29   & 31.55  & 35.88  & 44.10  & 44.36  & 36.31 & \makecell{40.21 \textcolor{red}{(\(\uparrow\)24.39\%)}}   \\
D+S+C+Dep   & \multicolumn{1}{c|}{Java}   & \cellcolor{gray!30}{\textbf{55.16}}   & 51.78    & 41.83    & 53.60   & 44.05  & 45.30  & 52.76  & 43.81  & 48.41 & \makecell{48.52 \textcolor{red}{(\(\uparrow\)50.08\%)}}   \\ \bottomrule
\end{tabular}}
\label{tab:code_generation}
\end{table*}

\begin{table*}[t]
\centering
\caption{Performance of LCMs in code completion across different contexts. Bold numbers on a gray background indicate the maximum values and red indicates the growth relative to the baseline.}
\scalebox{0.9}{
\begin{tabular}{@{}lcccccccccc|l@{}}
\toprule
Annotations & \multicolumn{1}{c|}{Language}                    & DSC-33B & DSC-6.7B & DSC-1.3B & SC2-15B & SC2-7B & SC2-3B & CL-34B & CL-13B & CL-7B & Average \\ \midrule
\multicolumn{12}{c}{BLEU}                                                                                                                 \\ \midrule
CS          & \multicolumn{1}{c|}{Python} & \cellcolor{gray!30}{\textbf{16.70}}   & 15.33    & 10.69    & 12.18   & 11.32  & 11.50  & 13.79  & 10.97  & 11.97 & 12.72   \\
CS+D        & \multicolumn{1}{c|}{Python} & \cellcolor{gray!30}{\textbf{23.08}}   & 21.21    & 14.92    & 17.68   & 14.11  & 13.35  & 15.95  & 17.04  & 15.16 & \makecell{16.94 \textcolor{red}{(\(\uparrow\)33.25\%)}}   \\
CS+D+C      & \multicolumn{1}{c|}{Python} & 28.87   & \cellcolor{gray!30}{\textbf{30.63}}    & 16.38    & 24.63   & 23.37  & 15.92  & 25.03  & 20.79  & 19.85 & \makecell{23.41 \textcolor{red}{(\(\uparrow\)84.12\%)}}   \\ \midrule
CS          & \multicolumn{1}{c|}{Java}   & \cellcolor{gray!30}{\textbf{31.86}}   & 30.81    & 20.53    & 21.67   & 16.82  & 15.91  & 22.36  & 21.28  & 26.09 & 23.04   \\
CS+D        & \multicolumn{1}{c|}{Java}   & \cellcolor{gray!30}{\textbf{40.58}}   & 39.14    & 31.71    & 31.21   & 15.45  & 19.16  & 33.31  & 29.00  & 32.83 & \makecell{30.27 \textcolor{red}{(\(\uparrow\)31.38\%)}}   \\
CS+D+C      & \multicolumn{1}{c|}{Java}   & \cellcolor{gray!30}{\textbf{51.96}}   & 51.11    & 40.56    & 38.89   & 30.09  & 29.90  & 49.51  & 29.37  & 45.25 & \makecell{40.88 \textcolor{red}{(\(\uparrow\)77.44\%)}}   \\ \midrule
\multicolumn{12}{c}{Edit Similarity}                                                                                                      \\ \midrule
CS          & \multicolumn{1}{c|}{Python} & \cellcolor{gray!30}{\textbf{29.18}}   & 28.45    & 21.77    & 22.89   & 19.47  & 19.91  & 25.48  & 22.79  & 22.99 & 23.66   \\
CS+D        & \multicolumn{1}{c|}{Python} & \cellcolor{gray!30}{\textbf{34.49}}   & 32.71    & 24.76    & 28.63   & 20.90  & 20.55  & 26.44  & 26.50  & 27.14 & \makecell{26.90 \textcolor{red}{(\(\uparrow\)13.71\%)}}   \\
CS+D+C      & \multicolumn{1}{c|}{Python} & 37.55   & \cellcolor{gray!30}{\textbf{38.59}}    & 25.04    & 33.09   & 29.64  & 21.98  & 31.40  & 29.85  & 25.96 & \makecell{30.73 \textcolor{red}{(\(\uparrow\)29.89\%)}}   \\ \midrule
CS          & \multicolumn{1}{c|}{Java}   & 33.09   & \cellcolor{gray!30}{\textbf{34.17}}    & 26.38    & 24.22   & 17.35  & 20.54  & 26.78  & 25.83  & 29.34 & 26.41   \\
CS+D        & \multicolumn{1}{c|}{Java}   & 40.30   & \cellcolor{gray!30}{\textbf{40.57}}    & 35.96    & 33.73   & 17.39  & 22.28  & 34.87  & 30.43  & 34.32 & \makecell{32.21 \textcolor{red}{(\(\uparrow\)21.94\%)}}   \\
CS+D+C      & \multicolumn{1}{c|}{Java}   & \cellcolor{gray!30}{\textbf{48.03}}   & 45.11    & 37.35    & 34.95   & 30.03  & 30.84  & 41.07  & 31.58  & 40.67 & \makecell{38.24 \textcolor{red}{(\(\uparrow\)44.80\%)}}   \\ \bottomrule
\end{tabular}}
\label{tab:code_completion}
\end{table*}
In this section, we explore the impact of context on the performance of LCMs in both code generation and code completion to reflect the necessity of context in complicated development scenarios.

\textbf{Code Generation.}
As shown in \autoref{tab:code_generation}, under the most basic context conditions (including only docstring and signature, referred to as D+S in the table), the average CodeBLEU scores for Python and Java are only 25.10 and 29.20, with average ES scores of 30.17 and 32.33, respectively. This indicates that the model struggles to achieve the desired effect in actual development with just basic information. After adding import information, the average CodeBLEU scores for Python and Java increase to 28.23 and 34.70, representing a 12.48\% and 18.83\% improvement, respectively, compared to using only basic information. The ES scores also increase by 8.04\% and 7.34\%, respectively. Furthermore, when file context containing import statements is added, the CodeBLEU scores increase by 21.88\% and 40.95\%, and the corresponding ES scores improve by 15.83\% and 24.39\%. Finally, after including function dependencies, the model's performance sees the greatest improvement, with CodeBLEU scores increasing by 31.90\% and 70.73\%. Additionally, we find that with more comprehensive contextual information, the performance of smaller models can surpass that of larger models. For example, Codellama-7B, when combined with full contextual information, achieves a CodeBLEU score of 49.53, while Codellama-34B's performance with basic conditions is only 27.54.

We attribute the performance improvement of LCMs to two main reasons: (1) \textit{Dependency information can enhance model performance}. Specifically, import information include third-party library dependencies imported by the code file, while file-level context includes not only third-party library dependencies but also function definitions within the code file (i.e., file-level dependencies). Function dependencies encompass these two types of dependency information as well as potential cross-file dependencies within the project. These three types of dependency information can significantly enhance the performance of LCMs in complex development scenarios. (2) \textit{Function background information can improve model performance}. Specifically, file context contains the logic and structure of the entire file's code, as well as background information on the function's purpose. This background information enables the model to understand the code task more comprehensively, thus improving performance.

\textbf{Code Completion.}
As shown in \autoref{tab:code_completion}, under basic context conditions, the average BLEU scores for Python and Java are only 12.72 and 23.04, with ES scores of 23.66 and 26.41, respectively. This indicates that merely providing code snippets (i.e., the preceding context of functions) makes it difficult for LCMs to infer the overall logic of the code. Therefore, after adding docstrings, the average BLEU scores for Python and Java increase by 33.25\% and 31.38\%, respectively, with ES scores improving by 13.71\% and 21.94\%. Furthermore, after adding file context, the average BLEU scores increase by 84.12\% and 77.44\%, with ES scores improving by 29.89\% and 44.80\%. These improvements also demonstrate that combining more contextual information can obviously enhance the performance of LCMs in code completion.

\finding{2}{Rich contextual information significantly enhances the performance of LCMs in complex development scenarios by incorporating function dependency information and background information. Furthermore, our experimental results indicate that the inclusion of extensive context can enable smaller models to outperform their larger counterparts.
}

\subsection{RQ3: Effect of Data Leakage on LCM Effectiveness}

\begin{table*}[ht]
\caption{Performance of LCMs in code generation and code completion on data from different time periods. Before and After indicate data before January 1, 2023 and data after September 15, 2023, respectively. Bold numbers indicate the difference between before and after.}
\scalebox{0.75}{
\begin{tabular}{@{}lc|ccc|ccc|ccc|ccc|ccc|ccc@{}}
\toprule
\multicolumn{1}{c}{\multirow{3}{*}{LCMs}} & \multicolumn{1}{c|}{\multirow{3}{*}{size}} & \multicolumn{9}{c|}{Code   Generation}                                                                                                                                                                                                                                                           & \multicolumn{9}{c}{Code   Completion}                                                                                                                                                                                                                                              \\ \cmidrule(l){3-20} 
\multicolumn{1}{c}{}                      & \multicolumn{1}{c|}{}                      & \multicolumn{3}{c|}{CodeBLEU}                                                                  & \multicolumn{3}{c|}{ES}                                                                        & \multicolumn{3}{c|}{EM}                                                                        & \multicolumn{3}{c|}{BLEU}                                                                       & \multicolumn{3}{c|}{ES}                                                                         & \multicolumn{3}{c}{EM}                                                         \\ \cmidrule(l){3-20} 
\multicolumn{1}{c}{}                      & \multicolumn{1}{c|}{}                      & \multicolumn{1}{c}{After} & \multicolumn{1}{c}{Before} & \multicolumn{1}{c|}{$\Delta$}                & \multicolumn{1}{c}{After} & \multicolumn{1}{c}{Before} & \multicolumn{1}{c|}{$\Delta$}                & \multicolumn{1}{c}{After} & \multicolumn{1}{c}{Before} & \multicolumn{1}{c|}{$\Delta$}                & \multicolumn{1}{c}{After} & \multicolumn{1}{c}{Before} & \multicolumn{1}{c|}{$\Delta$}                 & \multicolumn{1}{c}{After} & \multicolumn{1}{c}{Before} & \multicolumn{1}{c|}{$\Delta$}                 & \multicolumn{1}{c}{After} & \multicolumn{1}{c}{Before} & \multicolumn{1}{c}{$\Delta$} \\ \midrule
\multicolumn{20}{c}{Python}                                                                                                                                                                                                                                                                                                                                                                                                                                                                                                                                                                                                                                                    \\ \midrule
DSC                                       & \multicolumn{1}{l|}{1.3B}                  & 26.02                     & 29.96                      & \multicolumn{1}{l|}{\textbf{(+3.94)}} & 29.95                     & 33.27                      & \multicolumn{1}{l|}{\textbf{(+3.32)}} & 0.00                      & 1.19                       & \multicolumn{1}{l|}{\textbf{(+1.19)}} & 16.38                     & 22.81                      & \multicolumn{1}{l|}{\textbf{(+6.43)}}  & 25.04                     & 29.48                      & \multicolumn{1}{l|}{\textbf{(+4.44)}}  & 0.00                      & 1.19                       & \textbf{(+1.19)}      \\
DSC                                       & \multicolumn{1}{l|}{6.7B}                  & 32.01                     & 37.91                      & \multicolumn{1}{l|}{\textbf{(+5.90)}} & 36.54                     & 40.19                      & \multicolumn{1}{l|}{\textbf{(+3.65)}} & 0.00                      & 2.38                       & \multicolumn{1}{l|}{\textbf{(+2.38)}} & 30.63                     & 35.88                      & \multicolumn{1}{l|}{\textbf{(+5.25)}}  & 38.59                     & 42.07                      & \multicolumn{1}{l|}{\textbf{(+3.48)}}  & 0.00                      & 3.57                       & \textbf{(+3.57)}      \\
DSC                                       & \multicolumn{1}{l|}{33B}                   & 34.80                     & 34.59                      & \multicolumn{1}{l|}{\textbf{(-0.21)}} & 39.64                     & 39.88                      & \multicolumn{1}{l|}{\textbf{(+0.24)}} & 0.00                      & 1.19                       & \multicolumn{1}{l|}{\textbf{(+1.19)}} & 28.87                     & 32.69                      & \multicolumn{1}{l|}{\textbf{(+3.82)}}  & 37.55                     & 36.11                      & \multicolumn{1}{l|}{\textbf{(-1.44)}}  & 0.00                      & 0.00                       & \textbf{(0.00)}       \\
SC2                                       & \multicolumn{1}{l|}{3B}                    & 30.63                     & 32.70                      & \multicolumn{1}{l|}{\textbf{(+2.07)}} & 33.95                     & 34.57                      & \multicolumn{1}{l|}{\textbf{(+0.62)}} & 0.00                      & 1.19                       & \multicolumn{1}{l|}{\textbf{(+1.19)}} & 15.92                     & 22.56                      & \multicolumn{1}{l|}{\textbf{(+6.64)}}  & 21.98                     & 24.88                      & \multicolumn{1}{l|}{\textbf{(+2.90)}}  & 0.00                      & 0.00                       & \textbf{(0.00)}       \\
SC2                                       & \multicolumn{1}{l|}{7B}                    & 28.77                     & 31.48                      & \multicolumn{1}{l|}{\textbf{(+2.71)}} & 33.08                     & 32.98                      & \multicolumn{1}{l|}{\textbf{(-0.10)}} & 0.00                      & 1.19                       & \multicolumn{1}{l|}{\textbf{(+1.19)}} & 23.37                     & 29.13                      & \multicolumn{1}{l|}{\textbf{(+5.76)}}  & 29.64                     & 31.31                      & \multicolumn{1}{l|}{\textbf{(+1.67)}}  & 0.00                      & 0.00                       & \textbf{(0.00)}       \\
SC2                                       & \multicolumn{1}{l|}{15B}                   & 35.34                     & 35.54                      & \multicolumn{1}{l|}{\textbf{(+0.20)}} & 39.39                     & 40.74                      & \multicolumn{1}{l|}{\textbf{(+1.35)}} & 0.00                      & 0.00                       & \multicolumn{1}{l|}{\textbf{(0.00)}}  & 24.63                     & 32.17                      & \multicolumn{1}{l|}{\textbf{(+7.54)}}  & 33.09                     & 35.64                      & \multicolumn{1}{l|}{\textbf{(+2.55)}}  & 0.00                      & 0.00                       & \textbf{(0.00)}       \\
CL                                        & \multicolumn{1}{l|}{7B}                    & 29.72                     & 33.47                      & \multicolumn{1}{l|}{\textbf{(+3.75)}} & 34.10                     & 37.18                      & \multicolumn{1}{l|}{\textbf{(+3.08)}} & 0.00                      & 0.00                       & \multicolumn{1}{l|}{\textbf{(0.00)}}  & 19.85                     & 26.76                      & \multicolumn{1}{l|}{\textbf{(+6.91)}}  & 25.96                     & 30.40                      & \multicolumn{1}{l|}{\textbf{(+4.44)}}  & 0.00                      & 1.19                       & \textbf{(+1.19)}      \\
CL                                        & \multicolumn{1}{l|}{13B}                   & 26.49                     & 32.20                      & \multicolumn{1}{l|}{\textbf{(+5.71)}} & 31.98                     & 37.65                      & \multicolumn{1}{l|}{\textbf{(+5.67)}} & 0.00                      & 1.19                       & \multicolumn{1}{l|}{\textbf{(+1.19)}} & 20.79                     & 31.93                      & \multicolumn{1}{l|}{\textbf{(+11.14)}} & 29.85                     & 35.43                      & \multicolumn{1}{l|}{\textbf{(+5.58)}}  & 0.00                      & 1.19                       & \textbf{(+1.19)}      \\
CL                                        & \multicolumn{1}{l|}{34B}                   & 31.52                     & 35.36                      & \multicolumn{1}{l|}{\textbf{(+3.84)}} & 35.94                     & 41.51                      & \multicolumn{1}{l|}{\textbf{(+5.57)}} & 0.00                      & 1.19                       & \multicolumn{1}{l|}{\textbf{(+1.19)}} & 25.03                     & 37.37                      & \multicolumn{1}{l|}{\textbf{(+12.34)}} & 31.40                     & 40.10                      & \multicolumn{1}{l|}{\textbf{(+8.70)}}  & 0.00                      & 1.19                       & \textbf{(+1.19)}      \\ \midrule
Average                                   & \multicolumn{1}{l|}{-}                     & 30.59                     & 33.69                      & \multicolumn{1}{l|}{\textbf{(+3.10)}} & 34.95                     & 37.55                      & \multicolumn{1}{l|}{\textbf{(+2.60)}} & 0.00                      & 1.06                       & \multicolumn{1}{l|}{\textbf{(+1.06)}} & 22.83                     & 30.14                      & \multicolumn{1}{l|}{\textbf{(+7.31)}}  & 30.34                     & 33.94                      & \multicolumn{1}{l|}{\textbf{(+3.59)}}  & 0.00                      & 0.93                       & \textbf{(+0.93)}      \\ \midrule
\multicolumn{20}{c}{Java}                                                                                                                                                                                                                                                                                                                                                                                                                                                                                                                                                                                                                                                      \\ \midrule
DSC                                       & \multicolumn{1}{l|}{1.3B}                  & 36.94                     & 35.42                      & \multicolumn{1}{l|}{\textbf{(-1.52)}} & 36.04                     & 36.71                      & \multicolumn{1}{l|}{\textbf{(+0.67)}} & 0.00                      & 0.00                       & \multicolumn{1}{l|}{\textbf{(0.00)}}  & 40.56                     & 38.97                      & \multicolumn{1}{l|}{\textbf{(-1.59)}}  & 37.35                     & 38.65                      & \multicolumn{1}{l|}{\textbf{(+1.30)}}  & 0.00                      & 3.57                       & \textbf{(+3.57)}      \\
DSC                                       & \multicolumn{1}{l|}{6.7B}                  & 47.21                     & 43.13                      & \multicolumn{1}{l|}{\textbf{(-4.08)}} & 43.95                     & 43.08                      & \multicolumn{1}{l|}{\textbf{(-0.87)}} & 0.00                      & 0.00                       & \multicolumn{1}{l|}{\textbf{(0.00)}}  & 51.11                     & 52.35                      & \multicolumn{1}{l|}{\textbf{(+1.24)}}  & 45.11                     & 49.68                      & \multicolumn{1}{l|}{\textbf{(+4.57)}}  & 0.00                      & 2.38                       & \textbf{(+2.38)}      \\
DSC                                       & \multicolumn{1}{l|}{33B}                   & 46.90                     & 49.58                      & \multicolumn{1}{l|}{\textbf{(+2.68)}} & 45.41                     & 50.81                      & \multicolumn{1}{l|}{\textbf{(+5.40)}} & 0.00                      & 2.38                       & \multicolumn{1}{l|}{\textbf{(+2.38)}} & 51.96                     & 50.67                      & \multicolumn{1}{l|}{\textbf{(-1.29)}}  & 48.03                     & 48.44                      & \multicolumn{1}{l|}{\textbf{(+0.41)}}  & 0.00                      & 0.00                       & \textbf{(0.00)}       \\
SC2                                       & \multicolumn{1}{l|}{3B}                    & 35.72                     & 39.32                      & \multicolumn{1}{l|}{\textbf{(+3.60)}} & 35.88                     & 41.31                      & \multicolumn{1}{l|}{\textbf{(+5.43)}} & 0.00                      & 0.00                       & \multicolumn{1}{l|}{\textbf{(0.00)}}  & 29.90                     & 41.28                      & \multicolumn{1}{l|}{\textbf{(+11.38)}} & 30.84                     & 40.27                      & \multicolumn{1}{l|}{\textbf{(+9.43)}}  & 0.00                      & 5.95                       & \textbf{(+5.95)}      \\
SC2                                       & \multicolumn{1}{l|}{7B}                    & 30.83                     & 37.44                      & \multicolumn{1}{l|}{\textbf{(+6.61)}} & 31.55                     & 37.17                      & \multicolumn{1}{l|}{\textbf{(+5.62)}} & 0.00                      & 1.19                       & \multicolumn{1}{l|}{\textbf{(+1.19)}} & 30.09                     & 32.10                      & \multicolumn{1}{l|}{\textbf{(+2.01)}}  & 30.03                     & 31.12                      & \multicolumn{1}{l|}{\textbf{(+1.09)}}  & 1.19                      & 0.00                       & \textbf{(-1.19)}      \\
SC2                                       & \multicolumn{1}{l|}{15B}                   & 46.39                     & 45.53                      & \multicolumn{1}{l|}{\textbf{(-0.86)}} & 44.29                     & 46.50                      & \multicolumn{1}{l|}{\textbf{(+2.21)}} & 0.00                      & 0.00                       & \multicolumn{1}{l|}{\textbf{(0.00)}}  & 38.89                     & 41.81                      & \multicolumn{1}{l|}{\textbf{(+2.92)}}  & 34.95                     & 42.34                      & \multicolumn{1}{l|}{\textbf{(+7.39)}}  & 0.00                      & 0.00                       & \textbf{(0.00)}       \\
CL                                        & \multicolumn{1}{l|}{7B}                    & 36.45                     & 41.48                      & \multicolumn{1}{l|}{\textbf{(+5.03)}} & 36.31                     & 42.71                      & \multicolumn{1}{l|}{\textbf{(+6.40)}} & 0.00                      & 0.00                       & \multicolumn{1}{l|}{\textbf{(0.00)}}  & 45.25                     & 47.67                      & \multicolumn{1}{l|}{\textbf{(+2.42)}}  & 40.67                     & 42.36                      & \multicolumn{1}{l|}{\textbf{(+1.69)}}  & 0.00                      & 2.38                       & \textbf{(+2.38)}      \\
CL                                        & \multicolumn{1}{l|}{13B}                   & 43.90                     & 43.92                      & \multicolumn{1}{l|}{\textbf{(+0.02)}} & 44.36                     & 43.77                      & \multicolumn{1}{l|}{\textbf{(-0.59)}} & 0.00                      & 0.00                       & \multicolumn{1}{l|}{\textbf{(0.00)}}  & 29.37                     & 53.66                      & \multicolumn{1}{l|}{\textbf{(+24.29)}} & 31.58                     & 46.34                      & \multicolumn{1}{l|}{\textbf{(+14.76)}} & 0.00                      & 2.38                       & \textbf{(+2.38)}      \\
CL                                        & \multicolumn{1}{l|}{34B}                   & 46.07                     & 45.55                      & \multicolumn{1}{l|}{\textbf{(-0.52)}} & 44.10                     & 45.94                      & \multicolumn{1}{l|}{\textbf{(+1.84)}} & 0.00                      & 0.00                       & \multicolumn{1}{l|}{\textbf{(0.00)}}  & 49.51                     & 53.27                      & \multicolumn{1}{l|}{\textbf{(+3.76)}}  & 41.07                     & 46.45                      & \multicolumn{1}{l|}{\textbf{(+5.38)}}  & 0.00                      & 1.19                       & \textbf{(+1.19)}      \\ \midrule
Average                                   & \multicolumn{1}{l|}{-}                     & 41.16                     & 42.37                      & \multicolumn{1}{l|}{\textbf{(+1.22)}} & 40.21                     & 43.11                      & \multicolumn{1}{l|}{\textbf{(+2.90)}} & 0.00                      & 0.40                       & \multicolumn{1}{l|}{\textbf{(+0.40)}} & 40.74                     & 45.75                      & \multicolumn{1}{l|}{\textbf{(+5.02)}}  & 37.74                     & 42.85                      & \multicolumn{1}{l|}{\textbf{(+5.11)}}  & 0.13                      & 1.98                       & \textbf{(+1.85)}      \\ \bottomrule
\end{tabular}
}
\label{tab:time_performance}
\end{table*}
% \begin{figure*} [ht]
%     \centering
%     \includegraphics[width=\textwidth]{figures/data_leak.pdf}
%     \caption{Performance on Samples After September 15, 2023, and Before January 1, 2023}
%     \label{fig:data_leak}
% \end{figure*}

% \begin{figure*} [ht]
%   \begin{subfigure}{0.25\textwidth}
%     \includegraphics[width=\textwidth]{figures/data_leak_0.pdf}
%     \caption{Python Code Generation}
%   \end{subfigure}%
%   %\hspace{0.5cm}
%   \begin{subfigure}{0.25\textwidth}
%     \includegraphics[width=\textwidth]{figures/data_leak_1.pdf}
%     \caption{Java Code Generation}
%   \end{subfigure}%
%   %\hspace{0.5cm}
%   \begin{subfigure}{0.25\textwidth}
%     \includegraphics[width=\textwidth]{figures/data_leak_2.pdf}
%     \caption{Python Code Completion}
%   \end{subfigure}%
%     \begin{subfigure}{0.25\textwidth}
%     \includegraphics[width=\textwidth]{figures/data_leak_3.pdf}
%     \caption{Java Code Completion}
%   \end{subfigure}%
%   \caption{Performance on Samples After September 15, 2023, and Before January 1, 2023}
%   \label{fig:data_leak}
% \end{figure*}

To investigate the impact of data leakage on the performance of LCMs in code generation and code completion, we select data from two timestamps based on the knowledge cut-off dates of the LCMs (as shown in \autoref{tab:LCMs}): file created on or after September 15, 2023 (non-leaked data), and file created before January 1, 2023 (potentially leaked data). Based on the findings from RQ2, we employ code snippets, docstrings, and file context for code completion, while utilizing function signatures, docstrings, and file context for code generation in this section.
%we know that context improves the performance of LCMs. Therefore, in this phase, the code completion task leverages code snippets, docstrings, and file context, while the code generation task relies on function signatures, docstrings, and file context.

\textbf{Data leakage.} To more intuitively demonstrate the existence of data leakage, we introduce the Exact Match (EM) metric. Since each function in ComplexCodeEval contains many external dependencies and may also include variable definitions and other information, these external dependencies and variable information are unknown to LCMs during code generation. If LCMs have not seen the source code, it is almost impossible to generate code that exactly matches the reference code. However, in code completion, since the given code snippet may already include all external dependency calls and variable definitions, exact matches can occur.

As shown in \autoref{tab:time_performance}, in code generation, the EM ratio for all LCMs on non-leaked data is 0, while on leaked data, the average EM ratios for LCMs in Java and Python are 0.40\% and 1.06\%, respectively. In code completion, the EM ratios for LCMs on non-leaked data in Java and Python are 0.13\% and 0\%, respectively, while the corresponding ratios on leaked data are 1.98\% and 0.93\%. This finding reflects the existence of data leakage issues in LCMs.

\textbf{Impact of data leakage on LCMs' performance.} Based on \autoref{tab:time_performance}, we find that the CodeBLEU (BLEU) and ES scores of LCMs on leaked data are generally higher than the corresponding scores on non-leaked data, indicating that LCMs perform better on leaked data than on non-leaked data. Specifically, as shown in \autoref{tab:time_performance}, in code generation, the average CodeBLEU scores for LCMs on non-leaked data in Java and Python are 41.16 and 30.59, respectively, and the average ES scores are 40.21 and 34.95. On leaked data, the average CodeBLEU scores are 42.37 and 33.69, and the average ES scores are 43.11 and 37.55. Compared to non-leaked data, the average CodeBLEU scores on leaked data increased by 1.22 and 3.10, and the average ES scores increased by 2.90 and 2.60, respectively. In code completion, the average BLEU scores on leaked data and non-leaked data increased by 5.02 and 7.31, respectively, and the average ES scores increased by 5.11 and 3.59, respectively. These results suggest that data leakage may affect the performance of LCMs to some extent, leading to an overestimation of LCMs' performance. Additionally, since more information is provided to the model in code completion, the risk of overestimation is higher. The performance increase on leaked data compared to non-leaked data is also inconsistent across different models. For example, in Java code completion, the BLEU increase for StarCoder2-3B is 11.38, while for StarCoder2-7B it is only 2.01, indicating that using leaked data may not accurately assess the model's performance.

\finding{3}{LCMs exhibit superior performance on leaked data compared to non-leaked data, with the degree of performance improvement due to data leakage varying across different models. Consequently, evaluating LCMs' performance on code using leaked data may result in inaccurate and biased outcomes, failing to accurately reflect the models' true performance in real-world applications.
}

\section{Discussion}

\subsection{Performance with Different Docstrings} \label{sec:docstring_compare}
We use an LCM (Deep\-Seek-Coder-33B in this paper) to automatically generate docstrings for functions. We evaluate the LCMs using both the original docstrings and the automatically generated docstrings in this section. To avoid the influence of context on the results, we choose to evaluate Python code generation in the base context (i.e., including only the docstring and function signature). \autoref{tab:docstring_compare} shows the CodeBLEU and Edit Similarity (ES) scores for different LCMs when using original and automatically generated docstrings.

As seen in \autoref{tab:docstring_compare}, the CodeBLEU and ES scores of LCMs are generally higher when using automatically generated docstrings compared to the original docstrings. Specifically, by utilizing docstrings generated by LCM, the average CodeBLEU score of LCMs reaches 27.88, and the average ES score reaches 33.50, representing respective increases of 3.33 and 4.00.
This phenomenon can be attributed to two main reasons. First, automatically generated docstrings are typically more detailed and accurate than the original docstrings, providing the LCMs with comprehensive information about the functions that help the model better understand and generate code. Second, the quality of original docstrings varies widely, with many being relatively short and unclear, limiting the model's performance in code generation tasks. In contrast, automatically generated docstrings, which are carefully designed and iteratively optimized, can more accurately describe the target functions, considerably enhancing the model's performance.

\begin{table}[ht]
\centering
\caption{Effectiveness with two prompt versions of ComplexCodeEval. OD and GD indicate old docstring and generated docstring , respectively. Bold numbers indicate the difference between GD and OD.}
\begin{tabular}{@{}lc|ccc|ccc@{}}
\toprule
\multirow{2}{*}{LCMs} & \multirow{2}{*}{Size} & \multicolumn{3}{c|}{CodeBLEU} & \multicolumn{3}{c}{ES}        \\ \cmidrule(l){3-8} 
                      &                       & OD    & GD    & $\Delta$       & OD    & GD    & $\Delta$        \\ \midrule
DSC                   & 33B                   & 27.61 & 29.93 & \textbf{(+2.32)} & 31.51 & 34.75 & \textbf{(+3.24)} \\
DSC                   & 6.7B                  & 27.99 & 30.77 & \textbf{(+2.78)} & 33.40 & 35.92 & \textbf{(+2.52)} \\
DSC                   & 1.3B                  & 20.99 & 23.60 & \textbf{(+2.61)} & 25.92 & 28.26 & \textbf{(+2.34)} \\
SC2                   & 15B                   & 28.62 & 29.77 & \textbf{(+1.15)} & 33.72 & 36.32 & \textbf{(+2.60)} \\
SC2                   & 7B                    & 21.85 & 29.41 & \textbf{(+7.56)} & 26.56 & 35.26 & \textbf{(+8.70)} \\
SC2                   & 3B                    & 21.92 & 26.38 & \textbf{(+4.46)} & 27.18 & 31.22 & \textbf{(+4.04)} \\
CL                    & 34B                   & 27.89 & 30.69 & \textbf{(+2.80)} & 32.56 & 36.24 & \textbf{(+3.68)} \\
CL                    & 13B                   & 19.03 & 22.43 & \textbf{(+3.40)} & 24.56 & 29.53 & \textbf{(+4.97)} \\
CL                    & 7B                    & 25.07 & 27.98 & \textbf{(+2.91)} & 30.12 & 34.02 & \textbf{(+3.90)} \\ \midrule
Average               & -                     & 24.55 & 27.88 & \textbf{(+3.33)} & 29.50 & 33.50 & \textbf{(+4.00)} \\ \bottomrule
\end{tabular}
\label{tab:docstring_compare}
\end{table}

Overall, these results indicate that using high-quality docstrings is crucial for improving the performance of LCMs in code generation tasks. Future research can further explore how to optimize the process of generating docstrings to further enhance the performance of LCMs in code-related tasks.

\subsection{Performance With Different Benchmark}

\begin{table}[t]
\centering
\caption{Performance of Codellama-34B on different benchmark.}
\begin{tabular}{@{}cccc@{}}
\toprule
\multicolumn{1}{c|}{Metric}                           & Benchmark       & Python & Java  \\ \midrule
\multicolumn{4}{c}{Code Generation}                                                      \\ \midrule
\multicolumn{1}{c|}{\multirow{2}{*}{CodeBLEU}}        & CoderEval       & 26.38  & 35.88 \\
\multicolumn{1}{c|}{}                                 & ComplexCodeEval & 27.54  & 34.08 \\
\multicolumn{1}{c|}{\multirow{2}{*}{Edit Similarity}} & CoderEval       & 33.27  & 42.71 \\
\multicolumn{1}{c|}{}                                 & ComplexCodeEval & 32.93  & 36.93 \\ \midrule
\multicolumn{4}{c}{Code Completion}                                                      \\ \midrule
\multicolumn{1}{c|}{\multirow{2}{*}{BLEU}}            & CrossCodeEval   & 39.98  & 40.51 \\
\multicolumn{1}{c|}{}                                 & ComplexCodeEval & 13.79  & 22.36 \\
\multicolumn{1}{c|}{\multirow{2}{*}{Edit Similarity}} & CrossCodeEval   & 62.76  & 62.84 \\
\multicolumn{1}{c|}{}                                 & ComplexCodeEval & 25.48  & 26.78 \\ \midrule
\multicolumn{4}{c}{Test Case Generation}                                                 \\ \midrule
\multicolumn{1}{c|}{\multirow{2}{*}{CodeBLEU}}        & Method2Test     & -      & 35.79 \\
\multicolumn{1}{c|}{}                                 & ComplexCodeEval & -      & 29.90 \\
\multicolumn{1}{c|}{\multirow{2}{*}{Edit Similarity}} & Method2Test     & -      & 30.15 \\
\multicolumn{1}{c|}{}                                 & ComplexCodeEval & -      & 29.39 \\ \midrule
\multicolumn{4}{c}{API Recommendation}                                                   \\ \midrule
\multicolumn{1}{c|}{\multirow{2}{*}{F1}}              & APIbench        & 45.38  & 29.54 \\
\multicolumn{1}{c|}{}                                 & ComplexCodeEval & 52.24  & 38.92 \\
\multicolumn{1}{c|}{\multirow{2}{*}{Recall}}          & APIbench        & 45.84  & 29.72 \\
\multicolumn{1}{c|}{}                                 & ComplexCodeEval & 52.14  & 38.73 \\ \bottomrule
\end{tabular}
\label{tab:benchmark_compare}
\end{table}

In Section \ref{sec:RQ1}, the experimental results demonstrate that Codellama-34B achieves the best performance across four tasks. Consequently, this section focuses on Codellama-34B as the subject of our experiments. As illustrated in \autoref{tab:benchmark_compare}, we compare performance in code generation using CoderEval \cite{yu2024codereval}. Due to the high dependency on samples in CoderEval, Codellama-34B performs suboptimally on both CoderEval and ComplexCodeEval. In code completion, Codellama-34B performs better on CrossCodeEval \cite{ding2023crosscodeeval}, which focuses on simpler line-level completions, than on ComplexCodeEval. For test case generation, Codellama-34B exhibits slightly superior performance on Method2Test \cite{tufano2020unit} compared to ComplexCodeEval, whereas in API recommendation, Codellama-34B performs better on ComplexCodeEval than on APIbench \cite{peng2021revisiting}.

These results highlight the differences between ComplexCodeEval and various benchmarks associated with different tasks. Our benchmark not only encompasses multiple tasks but also illustrates performance disparities with existing benchmarks, thereby validating the necessity of introducing ComplexCodeEval.

\subsection{Threats To Validity}

\textbf{Threats in Benchmark Construction.} One potential threat is the way how we record the timestamps of the selected repositories. In order to mitigate the risk of not obtaining the corresponding time information from git commit records, we adopt a fallback mechanism: when the file creation time is unavailable, we use the project creation time as the file creation time; when the update time is unavailable, we use the creation time as the update time. This mechanism ensures the reliability and robustness of the samples. %This fallback mechanism may unintentionally introduce some level of bias when examining the effect of data leakage. %Another threat is that there may be room for improvement in the automatically generated docstrings. More advanced generation strategies (e.g., controlled text generation \cite{dekoninck2023controlled}) could be used to improve the quality of the docstrings. 

\textbf{Threats in Empirical Study.} Due to computing resource constraints, we do not conduct experiments on more open-source models (e.g., CodeGeex \cite{zheng2023codegeex}, WizardCoder \cite{luo2023wizardcoder}) and closed-source models (e.g., Gemini \cite{gemini2023}). Thus we select nine open-source LCMs ranging from 1.3B to 34B parameters and one closed-source LCM for experimentation, thereby covering a wide spectrum of model complexity and capability.

\section{Conclusion}

In this paper, we present ComplexCodeEval, a novel benchmark designed to evaluate the performance of LCMs in complex development scenarios. Unlike existing benchmarks, ComplexCodeEval is more adaptable and leverages a rigorous automated pipeline for data collection. Our experimental results highlight the limitations of LCMs in handling complex development tasks, while demonstrating that providing rich contextual information can substantially improve their performance. Furthermore, by comparing LCMs performance across data from different time periods, we emphasize that the use of leaked data in evaluations can lead to inaccurate results and introduce bias.

% In this paper, we introduce ComplexCodeEval, a novel benchmark for evaluating the performance of LCMs in complex development scenarios. Our benchmark is more adaptive compared to existing ones, utilizing a rigorous automated pipeline for data collection. Our experimental results reveal the limitations of LCMs in complex development contexts, while also highlighting that rich contextual information can significantly enhance the performance of LCMs in code. Additionally, by comparing the performance of LCMs on data from different timestamps, we point out that using leaked data to evaluate the code performance of LCMs can lead to inaccurate results and biases.

% In this work, we introduce ComplexCodeEval, a novel benchmark designed to evaluate LCMs. By employing a rigorous automated pipeline for data collection, ComplexCodeEval draws samples from a diverse range of code repositories, encompassing multiple critical domains of real-world development. To flexibly mitigate the impact of data leakage on LCM evaluation, we augment each sample with temporal information and provide rich annotations tailored to various downstream tasks. Experimental results reveal both the strengths and weaknesses of LCMs within real code repositories. We envision that ComplexCodeEval will foster a deeper understanding of LCMs within real-world code repositories and propel advancements in the field.

\section*{Acknowledgment}
This research is supported by the National Natural Science Foundation of China under project (No. 62472126), Natural Science Foundation of Guangdong Province (Project No. 2023A1515011959), Shenzhen-Hong Kong Jointly Funded Project (Category A, No. SGDX20230116091246007), Shenzhen Basic Research (General Project No. JCYJ20220531095214031), Shenzhen International Science and Technology Cooperation Project (No. GJHZ20220913143008015).

% \begin{acks}
% \end{acks}

\bibliographystyle{ACM-Reference-Format}
\bibliography{ref}

% \appendix
% \section{Prompt template}
%     \label{appendix-A}
%     \input{appendix/0_prompt-template}

% \section{ComplexCodeEval Details}
%     \label{appendix-B}
%     \input{appendix/1_ComplexCodeEval-details}

\end{document}